\documentstyle[12pt]{article}
\input epsf
\parskip=\medskipamount
\def\caH{{\cal H}}
\def\al{\alpha}

\def\la{\lambda}  \def\La{\Lambda}
\def\te{\theta}   
\def\om{\omega}   \def\Om{\Omega}
\def\k{\kappa}
   
\def\IC{\relax{\rm l\kern-.50 em C}}
\def\IE{\relax{\rm l\kern-.16 em E}}
\def\IK{\relax{\rm l\kern-.18 em K}}
\def\IL{\relax{\rm I\kern-.18 em L}}
\def\IN{\relax{\rm I\kern-.18 em N}}
\def\IR{\relax{\rm I\kern-.18 em R}}

\font\tenfrak=eufm10  \font\sevenfrak=eufm7  \font\fivefrak=eufm5
\newfam\frakfam
\textfont\frakfam=\tenfrak
\scriptfont\frakfam=\sevenfrak
\scriptscriptfont\frakfam=\fivefrak

\newtheorem{proposicion}{Proposition}

\def\wt{\widetilde}
\def\frac#1#2{{#1\over #2}}
\def\fracpd#1#2{\frac{\partial #1}{\partial #2}}

\def\bb{\vrule height5pt width5pt depth2pt}
\def\Cos{\mathop{\rm C}\nolimits}    
\def\Sin{\mathop{\rm S}\nolimits}    
\def\Tan{\mathop{\rm T}\nolimits}    
\def\k{\kappa}                       

\begin{document}

\title{ The quantum harmonic oscillator on the sphere \\ and the
hyperbolic plane }

\author{
Jos\'e F. Cari\~nena$\dagger\,^{a)}$,
Manuel F. Ra\~nada$\dagger\,^{b)}$,
Mariano Santander$\ddagger\,^{c)}$ \\
$\dagger$
    {\sl Departamento de F\'{\i}sica Te\'orica, Facultad de Ciencias} \\
    {\sl Universidad de Zaragoza, 50009 Zaragoza, Spain}  \\
$\ddagger$
    {\sl Departamento de F\'{\i}sica Te\'orica, Facultad de Ciencias} \\
    {\sl Universidad de Valladolid,  47011 Valladolid, Spain}
}

\maketitle
\date
\begin{abstract}
A nonlinear model of the quantum harmonic oscillator on
two-dimen\-sional spaces of constant curvature is exactly solved.
This model depends of a parameter $\la$ that is related with the
curvature of the space. Firstly the relation with other approaches
is discussed and then the classical system is quantized by
analyzing the symmetries of the metric (Killing vectors),
obtaining a $\la$-dependent invariant measure $d\mu_\la$ and
expressing the Hamiltonian as a function of the Noether momenta.
In the second part the quantum superintegrability of the
Hamiltonian and the multiple separability of the Schr\"odinger
equation is studied. Two $\la$-dependent Sturm-Liouville problems,
related with two different $\la$-deformations of the Hermite
equation, are obtained. This leads to the study of two
$\la$-dependent families of orthogonal polynomials both related
with the Hermite polynomials. Finally the wave functions
$\Psi_{m,n}$ and the energies $E_{m,n}$ of the bound states are
exactly obtained in both the sphere $S^2$ and the hyperbolic plane
$H^2$.
\end{abstract}

\begin{quote}
Position-dependent mass. Sch\-r\"o\-dinger equation.
Hermite-related equations. Hermite-related polynomials.
Sturm-Liouville problems.


Schr\"odinger, Dirac,
\end{quote}
{\vfill}

\footnoterule
{\noindent\small
$^{a)}${\it E-mail address:} {jfc@unizar.es}  \\
$^{b)}${\it E-mail address:} {mfran@unizar.es} \\
$^{c)}${\it E-mail address:} {msn@fta.uva.es}
\newpage

\section{Introduction }

   This article can be considered as a sequel or continuation of a
previous paper \cite{CRS06AnPh} which was devoted to the study of
a  quantum exactly solvable one-dimensional nonlinear oscillator with
quasi-harmonic behaviour. Now, our idea is to extend the results
and present a similar analysis but for the quantum version of the
two-dimensional nonlinear system. We follow the approach of
\cite{CRS06AnPh}, which contains the fundamental ideas and
motivation, and we will also make use of some properties
discussed, at the classical level, in \cite{CaRaSS04} (other
related papers are \cite{CaRaS04Romp}--\cite{CRS06Yer}).

  The following nonlinear differential equation
\begin{equation}
  (1 +\la x^2)\,\ddot{x} - \la x\,\dot{x}^2 + \al^2\,x  = 0\,,
  \quad(\rm \la\ a\ constant)\,, \label{Eq1}
\end{equation}
was study in Refs. \cite{MaLa74,LaRa03} as an example of a
nonlinear oscillator. In Lagrangian terms, the equation
(\ref{Eq1}) can be obtained from the following function
\begin{equation}
   L(x,v_x;\la)  = \frac{1}{2}\,\Bigl(\frac{v_x^2}{1 + \la\,x^2} \Bigr)
- \frac{\al^2}{2}\,\Bigl(\frac{x^2}{1 + \la\,x^2} \Bigr)\,,
\label{Lagn1}\end{equation}
  that clearly displays two very interesting characteristics:
the potential $V(\la)$ has a nonpolynomial character (this is not
a problem of an harmonic oscillator perturbed by higher order
terms of the form $\la x^{2m}$ with $m>1$), and the kinetic term
depends of a position-dependent mass.

  Let us briefly comment these two characteristics.

The form of the potential $V(\la)$ is shown in Figures I and II
for several values of $\la$ ($\la<0$ in Figure I and $\la>0$ in
Figure II). We see that for $\la<0$ the potential is a well with a
boundless wall at $x^2={1/|\la|}$ and for $\la>0$ we have that
$V\,{\to}\,(1/2)(\al^2/\la)$ for $x\,{\to}\,\pm\infty$. It can be
proved that
\begin{enumerate}
\item{} If $\la<0$ then the general solution of (\ref{Eq1}) is
given by
$$
  x = A \sin(\om\,t + \phi) \,,\quad
  \al^2 = (1 + \la\,A^2)\,\om^2 \,.
$$
\item{} If $\la>0$ then the general solution is given by
$$
  x = A \sin(\om\,t + \phi) \,,\quad
  \al^2 = (1 + \la\,A^2)\,\om^2 \,,
$$
when the energy $E$ is smaller than the value
$E_{\al,\la}=\al^2/(2\la)$, and by
$$
  x  = B \sinh(\Om t + \phi_1) \,,\quad
  \al^2 = (\la\,B^2 - 1)\,\om^2 \,,
$$
when the energy $E$ is greater than $E_{\al,\la}$.
\end{enumerate}

  The Schr\"odinger equation involving the potential
$\la(x^2/(1 + g x^2))$ has been studied by many authors making use
of different approaches \cite{BiDaS73}-\cite{Is02}. In some cases
the idea was to study the Hamiltonian $H= -\,d^2/dx^2 +x^2
+\la(x^2/(1 + g x^2))$ by applying perturbative, variational or
numerical techniques previously used for the system $H_m=
-\,d^2/dx^2+x^2 +\la x^{2m}$ with $m>1$. It is important to note
that in most of these papers the derivative part of the
Schr\"odinger equation was the standard one, that is, the equation
arising from a classical Hamiltonian with a quadratic term of the
form $(1/2)p^2$ and leading to a derivative term of the form
$-\,d^2/dx^2$, or to the corresponding two or three-dimensional
versions involving the Laplace operator in $\IE^2$ or $\IE^3$.
Nevertheless, we point out that Mathews and Lakshmanan studied in
Ref. \cite{MaLa75} the following quantum Hamiltonian
$$
  H = \frac{1}{2} \Bigl[\frac{1}{2} \bigl\{ p^2\,,\,(1-gx^2) \bigr\}_+
  + \frac{k\,x^2}{(1-gx^2)}\Bigr]\,,
$$
where the notation $\{A,B\}_+ = A B + B A$ is used.

  The second important point is the presence of a position-dependent
mass $m={(1+\lambda x^2)}^{-1}$ since, if the mass becomes a
spatial function, then the quantum version of the mass no longer
commutes with the momentum. Therefore, different forms of
presenting the kinetic term in the Hamiltonian $H$, that are
equivalent at the classical level, lead to different and
nonequivalent Schr\"odinger equations \cite{Le95}-\cite{JiYiJ05}.
This is an old question that remains as an important open problem
in the theory of quantization.

   In spite of these two characteristics, the quantum version
of this $\la$-dependent non-linear oscillator has been proved to be exactly
solvable \cite{CRS06AnPh}. The question of the order ambiguity in
the quantization of the Hamiltonian was solved by introducing a
prescription obtained from the analysis of the properties of the
classical system (existence of a Killing vector and a
$\la$-dependent invariant measure) and concerning the problems
arising from the nonpolynomial character of the potential, all of
them disappear when $V(\la)$ is studied with the appropriate
quantization of the position-dependent kinetic term.

On the other hand the following two-dimensional Lagrangian
$$
  L = \frac{1}{2}\,\Bigl(\frac{1}{1 + \la\,r^2} \Bigr)
  \Bigl[\,v_x^2 + v_y^2 + \la\,(x v_y - y v_x)^2 \,\Bigr]
  - \frac{\al^2}{2}\Bigl(\frac{r^2}{1 + \la\,r^2} \Bigr)\,,\quad
  r^2 = x^2+y^2\,,
$$
was proposed in Ref. \cite{CaRaSS04} as the appropriate
two-dimensional generalization, at the classical level,
of the Lagrangian (\ref{Lagn1}).
In fact the general solution of the Euler-Lagrange equations
which are given by
\begin{eqnarray}
  &&(1 + \la\,r^2)\,\ddot{x}  - \la\,\bigl[\, \dot{x}^2 + \dot{y}^2
  + \la\,(x\dot{y} - y\dot{x})^2\,\bigr]\,x +\al^2\,x = 0\,,\cr
  &&(1 + \la\,r^2)\,\ddot{y}  - \la\,\bigl[\, \dot{x}^2 + \dot{y}^2
  + \la\,(x\dot{y} - y\dot{x})^2\,\bigr]\,y +\al^2\,y = 0\,,
\nonumber\end{eqnarray}
is:
\begin{enumerate}
\item{} If $\la<0$ then the general solution is given by
$$
  x  = A \sin(\om t + \phi_1) \,,\quad
  y  = B \sin(\om t + \phi_2) \,,
$$
for all the values of the energy $E$. \item{} If $\la>0$ then the
general solution is given by
$$
  x  = A \sin(\om t + \phi_1) \,,\quad
  y  = B \sin(\om t + \phi_2) \,,
$$
when the energy $E$ is smaller than a certain value $E_{\al,\la}$,
and by
$$
  x  = A \sinh(\Om t + \phi_1) \,,\quad
  y  = B \sinh(\Om t + \phi_2) \,,
$$
when the energy $E$ is greater than this value, $E>E_{\al,\la}$.
\end{enumerate}
In both cases the coefficients $A$ and $B$ are related with the
coefficient $\al$ and the frequency $\om$ (oscillatory motions) or
with $\al$ and $\Om$ (unbounded motions). So we have
``quasi-harmonic" nonlinear oscillations in the case of bounded
motions and high energy scattering solutions when $\la>0$.
Moreover, the analysis of this nonlinear system proved the
existence of a relation with the linear harmonic oscillator on the
sphere $S^2$ or on the hyperbolic plane $H^2$ with the parameter
$\la$ playing the role of the (negative of the) curvature
$\kappa$.

  The main objective of this article is to quantize this
two-dimensional nonlinear oscillator as a deformation of the
harmonic oscillator in the sense that
\begin{enumerate}
\item{} All the fundamental properties of the linear system continue to
hold for $\la\ne 0$ but modified in a $\la$-dependent way.
\item{} The limit when $\la\to 0$ is well defined and when $\la=0$
all the characteristics of the quantum harmonic oscillator are
recovered.
\end{enumerate}
The idea is to prove that it is exactly solvable and to obtain the
energies and the corresponding wave functions.  Some of the main questions
to be discussed in this paper can be summarized in the following four
points:
\begin{itemize}

\item{} Relation of this $\la$-dependent nonlinear model with the harmonic
oscillator on spaces of constant curvature.

This quasi-harmonic nonlinear oscillator is related, at the
classical level, with the harmonic oscillator on the three spaces
of constant curvature ($S^2$, $\IR^2$, $H^2$). Now this relation
is considered for the quantized systems.

\item{} Analysis of the transition from the classical $\la$-dependent
system to the quantum one.

The two-dimensional $\la$-dependent kinetic term possesses three
Noether symmetries.   The main idea is to quantize the system by
using as Hilbert space the space $L^2(\IR,d\mu_\la)$ where
$d\mu_\la$ is a measure invariant under the Killing vectors
associated to the Noether symmetries.

\item{} Schr\"odinger separability and `quantum superintegrability'.

The two-dimensional Schr\"odinger equation is not separable in
$(x,y)$ coordinates because of the $\la$-dependent coupling
between the two degrees of freedom; nevertheless it is proved that
it admits separability in several coordinate systems. The
existence of this multiple separability is a property related with
`quantum superintegrability'.

\item{} Exact resolution of the $\la$-dependent Schr\"odinger
equation and families of $\la$-dependent orthogonal polynomials.

Two $\la$-dependent Sturm-Liouville problems related with two
different $\la$-deformations of the Hermite equation are obtained.
This leads to the study of two $\la$-dependent families of
orthogonal polynomials both related with the Hermite polynomials.

\end{itemize}

    In more detail, the plan of the article is as follows:
In Sec. 2 we study the relation of this nonlinear model with the harmonic
oscillator on spaces of constant curvature.
In Sec. 3 we first introduce a $\la$-dependent measure, we use it
for introducing a quantization rule and we obtain the
$\la$-dependent Schr\"odinger equation.
Sec. 4 is devoted to the analysis of the $\la$-dependent Schr\"odinger
separability and to solve two Hermite-related equations and
Sec. 5 to obtain the eigenfunctions $\Psi_{m,n}$ and energies $E_{m,n}$.
In Sec. 6 we briefly analyze the existence of another possible approach
and its relation with the presence of the angular momentum.
Finally, in Sec. 7 we make some final comments.

\section{On the relation of this nonlinear model with the harmonic
oscillator on spaces of constant curvature }

Although the first studies of the harmonic oscillator on curved
spaces are rather old  (the last chapter of Ref. \cite{Lieb}, that
is called `Nichteuklidische Mechanik', is devoted to the dynamics
on spaces with curvature; it first studies general properties and
then it consider the harmonic oscillator as a particular case; the
approach is mainly Newtonian), Ref. \cite{Hi79} is usually considered
as the more relevant paper for the modern approach to
this system (by modern we mean that it studies subjects such as
dynamical symmetries or quantum dynamics). We  recall that the
harmonic oscillator is a system that is well defined not only in
the Euclidean plane $\IE^2$ but also in the other two-dimensional
spaces of constant curvature, sphere $S^2$ and hyperbolic plane
$H^2$.

  In differential geometric terms, the three spaces with constant
curvature, sphere $S_{\k}^2$ ($\k>0$), Euclidean plane $\IE^2$,
and hyperbolic plane $H_{\k}^2$ ($\k<0$), can be considered as
three different situations inside a family of Riemannian manifolds
$M_{\k}^2=(S_{\k}^2,\IE^2,H_{\k}^2)$ with the curvature $\k$ as a
parameter $\k\in\IR$. In order to obtain mathematical expressions
valid for all the values of $\k$, it is convenient to
make use of the following $\kappa$-trigonometric functions
$$
  \Cos_{\k}(x) = \cases{
   \cos{\sqrt{\k}\,x}       &if $\k>0$, \cr
   {\quad}  1               &if $\k=0$, \cr
   \cosh\!{\sqrt{-\k}\,x}   &if $\k<0$, \cr}{\qquad}
   \Sin_{\k}(x) = \cases{
   \frac{1}{\sqrt{\k}} \sin{\sqrt{\k}\,x}     &if $\k>0$, \cr
   {\quad}   x                                &if $\k=0$, \cr
   \frac{1}{\sqrt{-\k}}\sinh\!{\sqrt{-\k}\,x} &if $\k<0$, \cr}
$$
then the expression of the differential element of distance in
geodesic polar coordinates $(R,\Phi)$ on the family
$M_{\k}^2=(S_{\k}^2,\IE^2,H_{\k}^2)$, can be written as follows
$$
  ds_{\k}^2 = dR^2 + \Sin_\k^2(R)\,d{\Phi}^2 \,,
$$
so it reduces to
$$
  ds_1^2 =    dR^2 + (\sin^2 R)\,d{\Phi}^2 \,,{\quad}
  ds_0^2 =    dR^2 + R^2\,d{\Phi}^2 \,,{\quad}
  ds_{-1}^2 = dR^2 + (\sinh^2 R)\,d{\Phi}^2\,,
$$
in the three particular cases of the unit sphere, the Euclidean plane,
and the `unit` Lobachewski plane. Note that $R$ denotes the distance
along a geodesic on the manifold $M_{\k}^2$; for example, in the
spherical $\k>0$ case, $R$ is the distance of the point to the origin
(e.g., the North pole) along a maximum circle.

If we make use of this formalism then the Lagrangian of the
harmonic oscillator on $M_{\k}^2$ is given by  \cite{RaSa02,RaSa03}
\begin{equation}
  \IL(\k) = (\frac{1}{2})\,\left(v_R^2 + \Sin_\k^2(R) v_\Phi^2\right)
    - (\frac{1}{2})\,\al^2\,\Tan_\k^2(R) . \label{Lagk}
\end{equation}
where the $\k$-dependent tangent is defined in the natural way
$\Tan_\k(R) = \Sin_{\k}(R)/\Cos_{\k}(R)$. In this way, the potential of the
harmonic oscillator on the unit sphere, on the Euclidean plane, or
on the unit Lobachewski plane, arise as the following three
particular cases
$$
  U_1(R) = (\frac{1}{2})\,\al^2\,\tan^2R     \,,{\quad}
  U_0(R) = (\frac{1}{2})\,\al^2\,R^2   \,,{\quad}
  U_{-1}(R) = (\frac{1}{2})\,\al^2\,\tanh^2R \,.
$$
The Euclidean oscillator $U_0(R)$ appears in this formalism as a
parabolic curve making a separation between two different
situations (see Figure III). Note also that in spherical $\k>0$
case, this Lagrangian describes in fact two oscillators with
centers in the north ($R=0$) and south ($\sqrt{\k}\,R=\pi$) poles
and with a boundary barrier in the equatorial circle.

Next we study the behaviour of $\IL(\k)$ under two different changes
of variables.

\begin{enumerate}

\item{} Let us consider the $\k$-dependent change $(R,\Phi) \to 
(r,\phi)$ given by
$$
  r  = \Sin_\k(R) \,,{\quad} \phi=\Phi\,,{\quad} \la=-\,\k\,.
$$
Then the Lagrangian $\IL(\k)$ becomes
$$
  L(\la) = \frac{1}{2}\,\Biggl(\frac{v_r^2}{1 + \la\,r^2} + 
r^2v_\phi^2 \,\Biggr)
  - \frac{\al^2}{2}\Bigl(\frac{r^2}{1 + \la\,r^2} \Bigr)\,.
$$
Therefore, if we change to Cartesian coordinates $(x,y)$ we arrive to
$$
  L(\la)  = \frac{1}{2}\,\Bigl(\frac{1}{1 + \la\,r^2} \Bigr)
  \Bigl[\,v_x^2 + v_y^2 + \la\,(x v_y - y v_x)^2 \,\Bigr]
  - \frac{\al^2}{2}\Bigl(\frac{r^2}{1 + \la\,r^2} \Bigr)\,,\quad
  r^2 = x^2+y^2\,,
$$
This function is just the Lagrangian obtained in Ref. \cite{CaRaSS04}
as the natural generalization of the one-dimensional Lagrangian $L(x,v_x;\la)$
for the nonlinear equation (\ref{Eq1}) of Mathews and Lakshmanan.

\item{} Let us now consider the $\k$-dependent change $(R,\Phi) \to (r',\phi)$
given by
$$
  r'  = \Tan_\k(R) \,,{\quad} \phi=\Phi\,.
$$
Then the Lagrangian $\IL(\k)$ becomes
$$
  L_H(\k) = \frac{1}{2}\,\Biggl(\frac{v_r'^2}{(1 + \k\,r'^2)^2} +
  \frac{r'^2 v_\phi^2}{(1 + \k\,r'^2)} \,\Biggr)
  - \frac{1}{2}\,\al^2 r'^2 \,,
$$
Therefore, if we change to Cartesian coordinates $(x,y)$ we arrive to
$$
  L_H(\k)  = \frac{1}{2}\, \frac{1}{(1 + \k \,r'^2)^2}
  \Bigl[\,v_x^2 + v_y^2 + \k \,(x v_y - y v_x)^2 \,\Bigr]
  - \frac{1}{2}\,\al^2 r'^2 \,,\quad
  r'^2 = x^2+y^2\,,
$$
This function is the Lagrangian studied by Higgs in Ref.
\cite{Hi79} (the study of Higgs was originally limited to a
spherical geometry but the idea can be easily extended to the
hyperbolic space).

\end{enumerate}

We note that these two changes are correct and both radial
variables, $r$ and $r'$, are well defined. In the hyperbolic
$\k<0$ case the two functions $\Sin_\k(R)$ and $\Tan_\k(R)$ are
positive for $R>0$ and concerning the spherical $\k>0$ case this
property is also true because then $R$ is restricted to a bounded
interval.

  The situation can be summarized as follows.
We have obtained three alternative ways of describing the harmonic
oscillator on spaces of constant curvature: the original system
$\IL(\k)$ and the two other approaches, $L(\la)$ and $L_H(\k)$,
obtained from it. Of course, everyone of these three different
approaches, $\IL(\k)$, $L(\la)$ and $L_H(\k)$, has its own
characteristics and advantages.

   A two--dimensional manifold $M$ can be described by using different
coordinate systems. If we consider it as an imbedded submanifold of $\IR^3$,
then the points of $M$ can be characterized by the three external coordinates,
as e.g. $(x,y,z)$, plus an additional constraint.
Nevertheless, in differential geometric terms, a more appropriate
approach is to develop the study by using two--dimensional systems
of coordinates intrinsically defined in $M$ (and without make reference
to the external space).
The Lagrangian $\IL(\k)$ is directly defined on the manifold
$M_{\k}^2=(S_{\k}^2,\IE^2,H_{\k}^2)$ and it uses the expression
of the differential element of distance $ds_{\k}^2$ in
geodesic polar coordinates $(R,\Phi)$ (see the Appendix).
Therefore, in differential geometric terms, this approach can be
considered as more formally correct than the other two.

  The Higgs approach \cite{Hi79,Le79} consider the motion on $S^n$,
embedded in the Euclidean space $\IE^{n+1}$, by means of a central
(also known as gnomonic) projection on a plane $\Pi^n$ tangent to
$S^n$ at a chosen point. This particular formalism leads to a
dynamics that is described, when  $n=2$, by the Lagrangian
$L_H(\k)$. This approach is very interesting because it states a
direct relation between the motion on a curved space, the sphere
$S^n$, and the motion on a plane. In fact, it has been studied by
many authors (see e.g. \cite{LeFe83}--\cite{HeBa06Sigma} and
references therein) mainly in relation of the theory of dynamical
symmetries.

The $\la$-dependent  Lagrangian $L(\la)$ has a certain similarity
with the Lagrangian of Higgs but nevertheless it does not coincide
with it: in the model of Higgs $\k$ (or $\la$) is present in the
kinetic term $T$ in a different way and the potential $V$ appears
as $\k$-independent; we will see that this affects, via the
Legendre transformation, to the Hamiltonian formalism. As we have
seen in Sec. I (Introduction) one of the advantages of this
approach is that the Euler-Lagrange equations can be directly
solved and the general solution has a rather simple form that can
be interpreted as ``quasi-harmonic" nonlinear oscillations. It is
clear that this $\la$-dependent formalism seems very appropriate
for solving equations or for other related calculus.

In what follows we will focuss our attention on the Hamiltonian dynamics
determined by the $\la$-dependent Lagrangian $L(\la)$.

\section{Quantization and $\la$-dependent Schr\"odinger equation }

Let us start our study considering the following Lagrangian
\begin{equation}
  L(\la)  = \frac{1}{2}\,\Bigl(\frac{1}{1 + \la\,r^2} \Bigr)
  \Bigl[\,v_x^2 + v_y^2 + \la\,(x v_y - y v_x)^2 \,\Bigr]
  - \frac{\al^2}{2}\Bigl(\frac{r^2}{1 + \la\,r^2} \Bigr)\,,\quad
  r^2 = x^2+y^2\,,
\label{Lagn2}\end{equation}
where the parameter $\la$ can take both
positive and negative values; of course it is clear that for
$\la<0$, $\la=-\,|\la|$, the function (and the associated
dynamics) will have a singularity at $1 -\,|\la|\,r^2=0$ and   we
shall restrict the study of the dynamics to the interior of the
interval $r^2<1/|\la|$ where the kinetic energy function
is positive definite.

  The Legendre transformation is
given by
$$
  p_x = \frac{v_x - \la\,J y}{\,1 + \la\,r^2\,} \,,\quad
  p_y = \frac{v_y + \la\,J x}{\,1 + \la\,r^2\,} \,,
$$
and the expression of the $\la$-dependent Hamiltonian turns out to be
\begin{equation}
  H(\la) = \frac{1}{2}\,\Bigl[\,p_x^2 + p_y^2
  + \la\,(x p_x + y p_y)^2 \Bigr]
  + \frac{\al^2}{2}\,\Bigl(\frac{r^2}{1 + \la\,r^2} \Bigr) \,.
\label{Hclas}\end{equation}

The transition from the classical system to the quantum one is a
difficult problem because of the ambiguities in the order of
positions and momenta. In Refs. \cite{CRS06AnPh,CaRaS04Romp} the
one-dimensional nonlinear oscillator was quantized by using a
prescription obtained from the existence of a one-dimensional
Killing vector $X(\la)$ and a $\la$-dependent measure $d\mu_\la$
in $\IR$ preserved by $X(\la)$. Next we prove that this approach,
that was successful for the one-dimensional system, admits a
direct generalization to this more difficult two-dimensional case.
We must begin with an analysis of the symmetries of the kinetic
energy term.

It was shown in \cite{CaRaSS04} that function $T(\la)$ representing 
the kinetic energy
$$
T(\la) = \frac{1}{2}\,\Bigl(\frac{1}{1 + \la\,r^2} \Bigr)
  \Bigl[\,v_x^2 + v_y^2 + \la\,(x v_y - y v_x)^2 \,\Bigr]
$$
is invariant under the action of the vector fields $X_1(\la)$,
$X_2(\la)$, and $X_J$, given by
\begin{eqnarray}
   X_1(\la) &=& \sqrt{\,1+\la\,r^2\,}\,\,\fracpd{}{x}  \,,\cr
   X_2(\la) &=& \sqrt{\,1+\la\,r^2\,}\,\,\fracpd{}{y}  \,,\cr
   X_J   &=& x\,\fracpd{}{y} - y\,\fracpd{}{x}\,,   {\nonumber}
\end{eqnarray}
in the sense that, if we denote by $X_r^t$, $r=1,2,J$, the natural lift to the
tangent bundle (phase space $\IR^2{\times}\IR^2$) of the vector field $X_r$,
\begin{eqnarray}
   X_1^t(\la) &=& \sqrt{\,1+\la\,r^2\,}\,\,\fracpd{}{x}
   +  \la\,\Bigl(\frac{x v_x + y 
v_y}{\sqrt{1+\la\,r^2\,}}\Bigr)\fracpd{}{v_x} \,,\cr
   X_2^t(\la) &=& \sqrt{\,1+\la\,r^2\,}\,\,\fracpd{}{y}
   +  \la\,\Bigl(\frac{x v_x + y 
v_y}{\sqrt{1+\la\,r^2\,}}\Bigr)\fracpd{}{v_y} \,,\cr
    X_J^t &=& x\,\fracpd{}{y} - y\,\fracpd{}{x}
   +  v_x\,\fracpd{}{v_y} - v_y\,\fracpd{}{v_x} \,,
{\nonumber}\end{eqnarray}
then the Lie derivatives of $T(\la)$ with respect to $X_r^t(\la)$
vanish, that is
$$
  X_r^t(\la)\Bigl(T(\la)\Bigr) = 0
  \,,\quad  r=1,2,J.
$$

  It is known that a symmetric bilinear form in the velocities $(v_x,v_y)$
can be considered as associated to a two-dimensional metric $ds^2$
in $\IR^2$. In this particular case, the function $T(\la)$
considered as a bilinear form determines the following
$\la$-dependent metric
\begin{equation}
  ds^2(\la) = \Bigl(\frac{1}{1 + \la\,r^2}\Bigr)\,
  \Bigl[\,(1 + \la\,y^2)\,dx^2 + (1 + \la\,x^2)\,dy^2 - 2 \la\,x y \,dx
\,dy\,\Bigr]\,.
\label{ds2}\end{equation}
Thus, in differential geometric terms, the three vector fields
$X_1(\la)$, $X_2(\la)$, and $X_J$, must be considered as three
Killing vector fields (infinitesimal generators of isometries) of
$ds^2(\la)$. These three symmetries of the kinetic term determine
three associate Noether momenta (a Noether momentum is a constant
of motion for the geodesic motion) given by
$$
  P_1(\la)= \frac{v_x - \la\,J y}{\sqrt{\,1 + \la\,r^2\,}}\,,\qquad
  P_2(\la)= \frac{v_y + \la\,J x}{\sqrt{\,1 + \la\,r^2\,}}\,,\qquad
  J = x v_y - y v_x\,,
$$
that become
$$
  P_1(\la)= \sqrt{\,1 + \la\,r^2\,} \,p_x\,,\qquad
  P_2(\la)= \sqrt{\,1 + \la\,r^2\,} \,p_y\,,\qquad
  J = x p_y - y p_x\,,
$$
in the Hamiltonian formalism. We note that the form of the angular
momentum $J$ is preserved by the Legendre map, in the sense that
we have $x p_y - y p_x=x v_y - y v_x$
(this is another one of the differences with the Higgs model).

\begin{proposicion}
The only  measure on the space $\IR^2$ that is invariant under the action of
the three vector fields $X_1(\la)$, $X_2(\la)$, and $X_J$, is given by
$$
   d\mu_\la = \Bigl(\frac{1}{\sqrt{1+\la\,r^2}}\Bigr)\,dx\,dy \,,
$$
up to a constant factor.
\end{proposicion}
{\it Proof:} The most general expression for a volume two-form on
the space $\IR^2$ is given by
$$
  \omega = \rho(x,y) \,dx \wedge dy
$$
Then the Lie derivatives of $\omega$ under $X_1(\la)$, $X_2(\la)$, and $X_J$,
are given by
\begin{eqnarray}
   {\cal L}_{X_1} \,\omega &=& 
\Bigl(\sqrt{1+\la\,r^2}\,\,\fracpd{\rho}{x}\Bigr)\,dx \wedge dy
  + \rho\, (d\,\sqrt{1+\la\,r^2}) \wedge dy            \cr
   {\cal L}_{X_2} \,\omega &=& 
\Bigl(\sqrt{1+\la\,r^2}\,\,\fracpd{\rho}{y}\Bigr)\,dx \wedge dy
  + dx  \wedge \rho\, (d\,\sqrt{1+\la\,r^2})            \cr
   {\cal L}_{X_J} \,\omega &=&  \Bigl(x\,\fracpd{\rho}{y} - 
y\,\fracpd{\rho}{x}\Bigr)
  \,dx \wedge dy
\nonumber \end{eqnarray} The condition ${\cal L}_{X_J} d\mu_\la=0$
implies that $\rho(x,y)$ must be a function $f(r)$ of $r$. Then
the two other conditions, ${\cal L}_{X_1} d\mu_\la=0$ and ${\cal
L}_{X_2} d\mu_\la=0$, lead to (for $r\ne 0$):
$$
  \sqrt{1+\la\,r^2}\,\frac{1}{r}\, \frac{df}{dr}
  + \frac{\la\,f}{\sqrt{1+\la\,r^2}} = 0  \,,
$$
with general solution given by
$$
  f = \frac{k}{\sqrt{1+\la\,r^2}}
$$
where $k$ is an arbitrary numerical constant. {\hfill\bb}

We will consider this proposition as the fundamental point for the
study of transition from the classical system to the quantum one.
In fact, this property suggests us to work with functions
and linear operators defined on the space obtained by considering
the two-dimensional real plane $\IR^2$ endowed with the measure
$d\mu_\la$ given by
\begin{equation}
   d\mu_\la = \Bigl(\frac{1}{\sqrt{1+\la\,r^2}}\Bigr)\,dx\,dy \,.\label{dmu}
\end{equation}
This means, in the first place, that the operators $\widehat{P_x}$
an $\widehat{P_y}$ representing the quantum version of of the
Noether momenta momenta $P_1$ an $P_2$ must be self-adjoint not in
the standard space $L^2(\IR)$ but in the space
$L^2(\IR,d\mu_\la)$. If we assume
\begin{eqnarray}
  \widehat{P_x} &=& -\,i\,\hbar\,\sqrt{1 + \la\,r^2}\,\fracpd{}{x} \,,\cr
  \widehat{P_y} &=& -\,i\,\hbar\,\sqrt{1 + \la\,r^2}\,\fracpd{}{y} \,.
\nonumber \end{eqnarray} then we arrive to the following
correspondence
\begin{eqnarray}
  (1 + \la\,r^2)\,p_x^2 \ &\to&\ -\,\hbar^2\,
  \Bigl(\sqrt{1 + \la\,r^2}\,\fracpd{}{x}\Bigr)
  \Bigl(\sqrt{1 + \la\,r^2}\,\fracpd{}{x}\Bigr) \,,\cr
  (1 + \la\,r^2)\,p_y^2 \ &\to&\ -\,\hbar^2\,
  \Bigl(\sqrt{1 + \la\,r^2}\,\fracpd{}{y}\Bigr)
  \Bigl(\sqrt{1 + \la\,r^2}\,\fracpd{}{y}\Bigr) \,,
\nonumber \end{eqnarray}
as well as
$$
  J^2 \ \to\ -\,\hbar^2\,
  \Bigl(x\,\fracpd{}{y} - y \,\fracpd{}{x}\Bigr)
  \Bigl(x\,\fracpd{}{y} - y \,\fracpd{}{x}\Bigr) \,,
$$
in such a way that the quantum version of the classical Hamiltonian
\begin{equation}
  H = \bigl(\frac{1}{2 m}\bigr)\,\Bigl[\,p_x^2 + p_y^2
  + \la\,(x p_x + y p_y)^2 \Bigr]
  + \bigl(\frac{1}{2}\bigr)\,g\,\Bigl(\frac{r^2}{1 + \la\,r^2} \Bigr)  \,,
    \quad g = m\al^2  \,.   \label{HClas}
\end{equation}
that can be rewritten as follows
$$
  H =\bigl(\frac{1}{2 m}\bigr)\, \Bigl[ P_1^2 + P_2^2 - \la\,J^2 \Bigr]
  + \bigl(\frac{1}{2}\bigr)\,g\,\Bigl(\frac{r^2}{1 + \la\,r^2} \Bigr) \,,
$$
is
\begin{eqnarray}
  \widehat{H} &=& - \frac{\hbar^2}{2 m}\, \Bigl[\,
  (1 + \la\,r^2)\,\fracpd{^2}{x^2} + \la\,x\,\fracpd{}{x} \,\Bigr]
  - \frac{\hbar^2}{2 m}\, \Bigl[\,
  (1 + \la\,r^2)\,\fracpd{^2}{y^2} + \la\,y\,\fracpd{}{y} \,\Bigr]    \cr
  &+& \la\, \frac{\hbar^2}{2 m}\, \Bigl[\, x^2\,\fracpd{^2}{y^2}
  + y^2\,\fracpd{^2}{x^2} - 2 x y\,\fracpd{^2}{x \,\partial y}
  - x\,\fracpd{}{x} - y\,\fracpd{}{y}\,\Bigr] + \bigl(\frac{1}{2}\bigr)\,g
\Bigl(\frac{r^2}{1 + \la\,r^2}\Bigr) \,.\label{Hqxy}
\end{eqnarray}

The first important property of this Hamiltonian is that it
admits the following decomposition
$$
  \widehat{H} = \widehat{H_1} + \widehat{H_2} - \la\, \widehat{J}^2
$$
where the three partial operators $\widehat{H_1}$, $\widehat{H_2}$
y $\widehat{J}^2$ are given by
\begin{eqnarray}
  \widehat{H_1} &=& - \frac{\hbar^2}{2 m}\, \Bigl[\,
  (1 + \la\,r^2)\,\fracpd{^2}{x^2} + \la\,x\,\fracpd{}{x} \,\Bigr]
+ \bigl(\frac{1}{2}\bigr)\,g \Bigl(\frac{x^2}{1 + \la\,r^2}\Bigr)   \cr
  \widehat{H_2} &=& - \frac{\hbar^2}{2 m}\, \Bigl[\,
  (1 + \la\,r^2)\,\fracpd{^2}{y^2} + \la\,y\,\fracpd{}{y} \,\Bigr]
+ \bigl(\frac{1}{2}\bigr)\,g \Bigl(\frac{y^2}{1 + \la\,r^2}\Bigr)   \cr
\widehat{J}^2 &=& -\, \frac{\hbar^2}{2 m}\, \Bigl[\,
  x^2\,\fracpd{^2}{y^2} + y^2\,\fracpd{^2}{x^2} - 2 x y 
\,\fracpd{^2}{x \,\partial y}
- x\,\fracpd{}{x} - y \,\fracpd{}{y}\,\Bigr]
\nonumber \end{eqnarray}
in such a way that the the total Hamiltonian $\widehat{H}$ commutes,
for any value of the parameter $\la$,  with each one
of the three partial terms
$$
  \bigl[ \widehat{H}\,,\widehat{H_1} \bigr] = 0\,,{\hskip 10pt}
  \bigl[\widehat{H}\,,\widehat{H_2} \bigr] = 0\,,{\hskip 10pt}
  \bigl[ \widehat{H}\,,\widehat{J}^2 \bigr] = 0\,.
$$
The vanishing of these three commutators means that the
$\la$-dependent  Hamiltonian (\ref{Hqxy}) describes a quantum
superintegrable system \cite{LeVi95}-\cite{CaNeOl06}. This
property was analyzed at the classical level in Ref.
\cite{CaRaSS04}; now we see that the quantization rule we have
applied preserves the superintegrability.

  Now, if we consider the Sch\"rodinger equation
$$
  \widehat{H}\,\Psi = E\,\Psi\,,
$$
as we have the following property
$$
  \bigl[ \widehat{H_1}\,,\widehat{H_2} - \la\, \widehat{J}^2 \bigr] = 
0\,,{\hskip 10pt}
  \bigl[ \widehat{H_1} - \la\, \widehat{J}^2\,,\widehat{H_2} \bigr] = 
0\,,{\hskip 10pt}
  \bigl[ \widehat{H_1} + \widehat{H_2}\,, \widehat{J}^2 \bigr] =
  0\,,
$$
then, we have three different sets of compatible observables and therefore
three different ways of obtaining a Hilbert basis of common eigenstates.
\begin{enumerate}
\item{} The two operators $\widehat{H_1}$ and $\widehat{H_2} -
\la\, \widehat{J}^2$ are a (complete) set of commuting observables;
therefore they represent two quantities that can be simultaneously
measured. Thus, the first way of looking for $\Psi$ is as a
solution of the following two equations
$$
\widehat{H_1}\,\Psi = E_1\,\Psi  \,,{\hskip 10pt}
\bigl(\widehat{H_2} - \la\, \widehat{J}^2\bigr)\,\Psi =
E_{2j}\,\Psi  \,.
$$
In this case the total energy is given by $E = E_1 + E_{2j}$ and
the associated wave function can be denoted by $\Psi(E_1,
E_{2j})$.

\item{} The two operators $\widehat{H_1}- \la\, \widehat{J}^2$ and
$\widehat{H_2}$ are a (complete) set of commuting observables. Thus,
the second way of looking for $\Psi$ is as a solution of the
following two equations
$$
\bigl(\widehat{H_1} - \la\, \widehat{J}^2\bigr)\,\Psi =
E_{1j}\,\Psi  \,,{\hskip 10pt} \widehat{H_2}\,\Psi = E_2\,\Psi \,.
$$
In this case we have $E = E_{1j} + E_2$ and $\Psi$ can be denoted
by $\Psi(E_{1j}, E_2)$.

\item{} The third (complete) set of commuting observables is
provided by $\widehat{H_1} + \widehat{H_2}$ and $\widehat{J}^2$.
So in this case we have
$$
  \bigl(\widehat{H_1} + \widehat{H_2}\bigr)\,\Psi = E_{12}\,\Psi 
\,,{\hskip 10pt}
  \widehat{J}\,\Psi = j\,\Psi   \,.
$$
Thus, the two physically measurable quantities are $E_{12}$ and
the angular momentum $j$, the total energy is given by $E = E_{12}
- \la\,j^2$ and the wave function so defined can be denoted by 
$\Psi(E_{12},j)$.
\end{enumerate}

The existence of these three alternative descriptions arises
from the presence of the term $\la\,\widehat{J}^2$ inside the
kinetic part of the Hamiltonian. Notice that the second approach
is just the symmetric of the first one but, although they
are closely related, they lead however to different solutions with
different properties; that is, $\Psi(E_1, E_{2j}) \ne \Psi(E_{1j},
E_2)$. This fact is a consequence of the nonlinear character of
the model since in the linear limit, when $\la\to 0$, then both
descriptions coincide.

  Let us consider the following quantum Hamiltonian
\begin{eqnarray}
  \widehat{H} &=& - \frac{\hbar^2}{2 m}\, \Bigl[
  (1 + \la\,r^2)\,\fracpd{^2}{x^2} + \la\,x\,\fracpd{}{x} \,\Bigr]
  - \frac{\hbar^2}{2 m}\, \Bigl[
  (1 + \la\,r^2)\,\fracpd{^2}{y^2} + \la\,y\,\fracpd{}{y} \,\Bigr]
  + \la\, \frac{\hbar^2}{2 m}\, \Bigl[\, x^2\,\fracpd{^2}{y^2} \cr
  && + y^2\,\fracpd{^2}{x^2} - 2 x y \,\fracpd{^2}{x \,\partial y}
  - x\,\fracpd{}{x} - y \,\fracpd{}{y}\,\Bigr]
  + \bigl(\frac{1}{2}\bigr)\,m\al\,(\al + \frac{\hbar}{m}\,\la)
  \Bigl(\frac{r^2}{1 + \la\,r^2}\Bigr)
\label{Hqxyv2}\end{eqnarray} where we have slightly modified the
value of the parameter $g$ that now is given by $g = m\al^2 +
\la\,\hbar\al$ (this is done to coincide with the notation of the
one-dimensional system in \cite{CRS06AnPh}). It is also convenient
to simplify this function $\widehat{H}$ by introducing
adimensional variables $(\tilde{x},\tilde{y},\La,e)$ defined by
$$
  x = \Bigl(\sqrt{\frac{\hbar}{m\al}}\,\Bigr)\,\tilde{x} \,,\quad
  y = \Bigl(\sqrt{\frac{\hbar}{m\al}}\,\Bigr)\,\tilde{y} \,,\quad
  \la = \Bigl(\frac{m\al}{\hbar}\Bigl)\,\La  \,,\quad
  E = (\hbar\al)\,e  \,,\
$$
in such a way that the following  relation holds
$$
  1 + \la\,r^2 = 1 + \La\,\tilde{r}^2 \,,\quad
  \tilde{r}^2 = \tilde{x}^2+\tilde{y}^2  \,.
$$
The Schr\"odinger equation takes then the following form
\begin{eqnarray}
   && - \frac{1}{2}\, \Bigl[\, (1 + \La\tilde{r}^2)\,\fracpd{^2}{\tilde{x}^2}
  + \La\,\tilde{x}\,\fracpd{}{\tilde{x}} \,\Bigr]\,\Psi
  - \frac{1}{2}\, \Bigl[\, (1 + \La\,\tilde{r}^2)\,\fracpd{^2}{\tilde{y}^2}
  + \La\,\tilde{y}\,\fracpd{}{\tilde{y}} \,\Bigr]\,\Psi
  + \frac{\La}{2}\, \Bigl[\, \tilde{x}^2\,\fracpd{^2}{\tilde{y}^2}  \cr
  && +\, \tilde{y}^2\,\fracpd{^2}{\tilde{x}^2} -
  2 \tilde{x} \tilde{y} \,\fracpd{^2}{\tilde{x} \,\partial \tilde{y}}
- \tilde{x}\,\fracpd{}{\tilde{x}} - \tilde{y} 
\,\fracpd{}{\tilde{y}}\,\Bigr]\,\Psi +
\bigl(\frac{1}{2}\bigr)\,(1+\La) \Bigl(\frac{\tilde{r}^2}{1 +
\La\,\tilde{r}^2}\Bigr) \,\Psi = e\,\Psi
\label{EcSchxy}\end{eqnarray}

\section{Resolution of the Schr\"odinger equation I}

In the following all the variables, parameters and equations are
adimensional. Nevertheless, and for ease of notation, we will drop
the use of the tilde and write the variables just as $x$, $y$, $r$
and so on.

\subsection{Separability}

The Schr\"odinger equation (\ref{EcSchxy}) is not separable in
Cartesian $(x,y)$ coordinates because of the $\La$-dependent term.
Nevertheless, at this point we recall that it was proved in \cite{CaRaSS04}
that, at the classical level, the Hamilton-Jacobi equation
$$
   \Bigl(\fracpd{S}{x}\Bigr)^2 + \Bigl(\fracpd{S}{y}\Bigr)^2
  + \la\,\Bigl(x\,\fracpd{S}{x} + y\,\fracpd{S}{y}\Bigr)^2
  +  \frac{\al^2}{2}\,\Bigl(\frac{x^2+y^2}{1+\La\,(x^2+y^2)}\Bigr) = 2 E \,.
$$
admits separability in the following three
different orthogonal coordinate systems:
\begin{enumerate}
\item{} $\La$-dependent coordinates $(z_x,y)$ with $z_x$ defined by
$z_x = x/\sqrt{1 + \La\,y^2}$\,.

\item{} $\La$-dependent coordinates $(x,z_y)$ with $z_y$ defined by
$z_y = y/\sqrt{1 + \La\,x^2}$\,.

\item{} Polar coordinates $(r,\phi)$.
\end{enumerate}

The expression of the potential $V(\la)$ in these three systems is
\begin{eqnarray}
  V(\la) &=& \frac{1}{2}\,\Bigl(\frac{1}{1+\La\,y^2}\Bigr)
  \Bigl[\frac{z_x^2}{1+\La\,z_x^2}  +  y^2 \Bigr]      \cr
   &=&  \frac{1}{2}\,\Bigl(\frac{1}{1+\La\,x^2}\Bigr)
  \Bigl[ x^2  +  \frac{z_y^2}{1+\La\,z_y^2} \Bigr]    \cr
   &=&  \frac{1}{2}\,\Bigl(\frac{r^2}{1+\La\,r^2}\Bigr)\,.
\nonumber \end{eqnarray}

Since it is known the existence of a close relation between
(additive) classical Hamilton-Jacobi separability and
(multiplicative) quantum Sch\"rodinger separability
\cite{BeChRa02a,BeChRa02b} it seems natural to make use of these
three coordinate systems for the study of this quantum problem.

Next we start our study with the first coordinate system.

Using $(z_x,y)$ coordinates, the Schr\"odinger equation becomes
\begin{eqnarray}
  && - \frac{1}{2}\, \Bigl[\,
  \Bigl(\frac{1 + \La\,z_x^2}{1 + \La\,y^2}\Bigr)\,\fracpd{^2}{z_x^2}
  + \Bigl(\frac{\La\,z_x}{1 + \La\,y^2}\Bigr)\,\fracpd{}{z_x}  \,\Bigr]\,\Psi
  - \frac{1}{2}\, \Bigl[\,
  (1 + \La\,y^2)\,\fracpd{^2}{y^2} + 2\La\,y\,\fracpd{}{y} \,\Bigr]\,\Psi \cr
  &&+ \bigl(\frac{1}{2}\bigr)\,(1+\La) \Bigl(\frac{1}{1 + \La\,y^2}\Bigr)
\Bigl(\frac{z_x^2}{1 + \La\,z_x^2} + y^2\Bigr) \,\Psi = e\,\Psi
\,. \label{EcSchzy}\end{eqnarray}
We assume a solution in the form
$$
  \Psi(z,y)  = Z(z_x)\,Y(y) \,,
$$
where $Z$ and $Y$ are, respectively, functions of $z_x$ and $y$ alone.
Substituting this expression in the Eq. (\ref{EcSchzy}) we arrive to
\begin{eqnarray}
   && - \frac{1}{2}\,\frac{1}{Z}\, \Bigl[\,
  \bigl(1 + \la\,z_x^2\bigr)\,Z'' + \bigl(\la\,z_x\bigr)\,Z'  \,\Bigr]
+ \bigl(\frac{1}{2}\bigr)\,(1+\La)\,\Bigl(\frac{z_x^2}{1 +
\la\,z_x^2}\Bigr) \cr
  & =&  \frac{1}{2}\,\frac{1}{Y}\, \Bigl[\,
  (1 + \la\,y^2)^2\,Y'' + 2\la\,y\,(1 + \La\,y^2)\,Y'\,\Bigr]
- \bigl(\frac{1}{2}\bigr)\,(1+\La)\,y^2 + (1 + \La\,y^2)\,e
\nonumber\end{eqnarray}
that leads to the two following ordinary equations
\begin{eqnarray}
   && - \frac{1}{2}\,\frac{1}{Z}\, \Bigl[\, \bigl(1 + \La\,z_x^2\bigr)\,Z''
  + \bigl(\La\,z_x\bigr)\,Z'  \,\Bigr]
+ \bigl(\frac{1}{2}\bigr)\,(1+\La)\,\Bigl(\frac{z_x^2}{1 +
\La\,z_x^2}\Bigr) = \mu   \,,\cr
  && - \frac{1}{2}\,\frac{1}{Y}\,\Bigl[\,
  (1 + \La\,y^2)^2\,Y'' + 2\La\,y\,(1 + \La\,y^2)\,Y'\,\Bigr]
+ \bigl(\frac{1}{2}\bigr)\,(1+\La)\,y^2 - (1 + \La\,y^2)\,e =
-\,\mu  \,, \nonumber\end{eqnarray}
where  $\mu$  denotes the separation constant.

Consequently the $\La$-dependent Schr\"odinger equation is in fact
separable in the $(z_x,y)$ coordinates. Thus the two-dimensional
problem has been decoupled in two one-dimensional equations.

\subsection{Power series resolution of the $Z$-equation}

The first equation to be solved is
\begin{equation}
  \bigl(1 + \La\,z_x^2\bigr)\,Z'' + \bigl(\La\,z_x\bigr)\,Z'
  - (1+\La) \Bigl(\frac{z_x^2}{1 + \La\,z_x^2}\Bigr)\,Z  = -2 \mu \,Z
\label{EcZ}\end{equation}
This equation coincides (up to the appropriate changes of notation)
with the equation of the one-dimensional nonlinear oscillator studied
in Ref. \cite{CRS06AnPh}.
Therefore, we directly explain the characteristics of the solution.

Firstly, using the following factorization for the function $Z$
\begin{equation}
   Z(z_x,\La) = p(z_x,\La)\,(1 + \La\,z_x^2)^{-\,1/(2\La)} \label{Z(zxLa)}\,,
\end{equation}
the function $p=p(z_x,\La)$ must satisfy the differential
equation
\begin{equation}
   (1 + \La\,z_x^2) p'' + (\La - 2) z_x p' + (2 \mu -1) p = 0 \,,
\label{Ecpz}\end{equation}
that represents a $\La$-deformation of the Hermite equation.
Secondly, this new equation can be solved by the use of power
series expansions. Assuming
$$
  p(z_x,\La) = \sum_{n=0}^\infty\,c_n(\La)\,z_x^n
   = c_0(\La) + c_1(\La)\,z_x + c_2(\La)\,z_x^2 + \dots
$$
the following $\La$-dependent recursion relation is obtained
$$
  c_{n+2} = (-1)\,\frac{c_n}{(n+2)(n+1)}\,
  \Bigl[\,n\,(\La\,n -\,2) + (2 \mu -1)\,\Bigr]\,,\qquad  n=0,1,2,\ldots
$$
The general solution $p(z_x)$, defined in the interval $z_x^2 <
1/\La$, is given by the linear combination $p = c_0 p_1 + c_1 p_2$
where $p_1(z_x)$ and $p_2(z_x)$ are the two solutions determined by
$p_1(0)=1,\ p_1'(0)=0$ and $p_2(0)=0,\ p_2'(0)=1$, respectively.
If there exists a certain integer $m$ such that the coefficient
$\mu=\mu_m$ is given by
$$
  2 \mu_m = 2\,m + 1 - \La\,m^2 \,,
$$
then we have  $c_m\ne 0$, $c_{m+2} = 0$, and one of the two
solutions is a polynomial of order $m$.

The polynomial solutions are given by
\begin{itemize}
\item{} Even index (even power polynomials)
\begin{eqnarray}
  {\cal P}_{2p} &=& \sum_{r=0}^{r=p} \,c_{2r}\,z_x^{2r} \cr
  c_{2r} &=& (-1)^r\,\frac{a_0}{2r\,!}\,p'\,(p'-2)(p'-4)\dots(p'-2(r-1)) \cr
  &&{\hskip50pt}\bigl[\,2 - \La\,p'\,\bigr]\bigl[\,2 - \La (p'+2)\,\bigr]
  \bigl[\,2 - \La(p'+4)\,\bigr]\dots\bigl[\,2 - \La\,(p'+2(r-1))\,\bigr] \,,
\nonumber \end{eqnarray}
where we have introduced the notation $p'=2p$. More specifically,
the expressions of the first solution $p_1(z_x)$, in the
particular cases of $p'=0,2,4$, are given by:
\begin{eqnarray}
  {\cal P}_0 &=& 1 \,,\cr
  {\cal P}_2 &=& 1 - 2(1 - \La)z_x^2  \,,\cr
  {\cal P}_4 &=& 1 - 4(1 - 2\La)z_x^2 + (\frac{4}{3})(1 - 2\La)(1 - 3\La)z_x^4
- 4\La)z_x^4
\nonumber \end{eqnarray}

  \item{} Odd index  (odd power polynomials)
\begin{eqnarray}
   {\cal P}_{2p+1} &=& \sum_{r=0}^{r=p} \,a_{2r+1}\,z_x^{2r+1} \cr
   c_{2r+1} &=& 
(-1)^{r+1}\,\frac{a_1}{2r+1\,!}\,(p'-1)(p'-3)\dots(p'-(2r-1)) \,,\cr
   &&{\hskip80pt}\bigl[\,2 - \La\,(p'+1)\,\bigr]\bigl[\,2 - \La (p'+3)\,\bigr]
   \dots\bigl[\,2 - \La\,(p'+(2r-1))\,\bigr]\,,   \nonumber
\end{eqnarray}
where we have introduced the notation $p'=2p+1$. More
specifically, the expressions of the second solution $p_2(z_x)$
for $p'=1,3,5$, are given by:
\begin{eqnarray}
  {\cal P}_1 &=& z_x  \,,\cr
  {\cal P}_3 &=& z_x - (\frac{2}{3}) (1 - 2\La)z_x^3  \,,\cr
  {\cal P}_5 &=& z_x - (\frac{4}{3}) (1 - 3\La)z_x^3
   + (\frac{4}{15})(1 - 3\La)(1 - 4\La)z_x^5   \,.\nonumber
\end{eqnarray}
\end{itemize}

\subsection{Power series resolution of the $Y$-equation}

The second equation to be solved is
\begin{equation}
  (1 + \La\,y^2)\,Y'' + \bigl( 2\La\,y\bigr)\,Y'
  -\,(1+\La) \Bigl(\frac{y^2}{1 + \La\,y^2}\Bigr)\,Y
  + 2\,e\,Y = 2\,\Bigl(\frac{\mu}{1 + \La\,y^2}\Bigr)\,Y
\label{EcY}\end{equation}
that, although it has certain similarity with the Eq. (\ref{EcZ}),
it does not coincide with it (two differences: the coefficient
$2\La$ and the $\mu$-dependent right-hand term).
The main reason for this asymmetry is that, when introducing
separability in the Schr\"odinger equation, the angular momentum
term $\widehat{J}^2$ was displaced into this second equation.

We start our study with the following two steps.

Step 1. Introduction of a new quantum number

It is convenient to decompose the energy  $e$ as the following sum
$$
  e = \mu + \nu
$$
where $\nu$ is a new parameter. Then the equation (\ref{EcY})
transforms into
$$
  (1 + \La\,y^2)\,Y'' + \bigl(2\La\,y\bigr)\,Y'
  -\,(1+\La - 2 \La\,\mu)\,\Bigl(\frac{y^2}{1 + \La\,y^2}\Bigr)\,Y
+ 2\,\nu\,Y = 0 \,,
$$
which looks more similar to previous first equation for the function $Z$.

Step 2. Factorization of the function $Y$

Let us rewrite the previous equation as follows
$$
   (1 + \La\,y^2)\,Y'' + \bigl( 2\La\,y\bigr)\,Y'
  - G_\mu^2 \Bigl(\frac{y^2}{1 + \La\,y^2}\Bigr)\,Y + 2\,\nu\,Y = 0 \,,\quad
  G_\mu^2 = 1 +  (1 - 2 \mu) \La \,.
$$
Firstly, it can be verified that the function
$\Psi_{\infty}$ defined by
$$ \Psi_{\infty} = (1 + \La\,y^2)^{-\,G_\mu/(2\La)}
$$
satisfies the following property
$$
  \Biggl[\, (1 + \La\,y^2) \frac{d^2}{dy^2} + 2\La\, y\,\frac{d}{dy}
  -\,G_\mu^2 \Bigl(\frac{y^2}{1 + \La\,y^2}\Bigr)  \Biggr]\Psi_{\infty}
  = - G_\mu\,\Psi_{\infty}  \,.
$$
Thus, $\Psi_{\infty}$ is the exact solution of the Eq. (\ref{EcY}) in
the very particular case of $\nu=(1/2) G_\mu$ and can be
considered as representing, in the general case $2\,\nu \ne
G_\mu$, the asymptotic behaviour of the solution. Consequently,
this property suggests the following factorization
\begin{equation}
  Y(y,\La) = q(y,\La)\,(1 + \La\,y^2)^{-\,G_\mu/(2\La)} \,,
\end{equation}
and then the new function $q(y,\La)$ must satisfy the differential
\begin{equation}
  (1 + \La\,y^2) q'' + 2 (\La - G_\mu)y q' + (2 \nu - G_\mu) q = 0 \,,
\label{Ecqy}\end{equation}
that turns out to be a new $\La$-deformation of the Hermite equation.

Assuming a power expansion for the solution
$$
  q(y,\La) = \sum_{n=0}^\infty\,c_n(\La)\,y^n
   = c_0(\La) + c_1(\La)\,y + c_2(\La)\,y^2 + \dots
$$
the equation leads to
$$
  \sum_{n=0}^\infty\,\left[ (n+2)(n+1)\,c_{n+2}+ \La\,n (n-1)c_n\,y^n
   + 2(\La - G_\mu)\,n\,c_n + (2 \nu - G_\mu)\,c_n \right]\,y^n = 0  \,,
$$
and we obtain the following $\La$-dependent recursion relation
$$
  c_{n+2} = (-1)\,\frac{c_n}{(n+2)(n+1)}\,
         \Bigl[\,\La\,n\,(n -1) - G_\mu (2 n + 1) + 2 \nu \,\Bigr]
$$
Note that this relation shows that, as in the $\La=0$ case, even
power coefficients are related among themselves and the same is
true for odd power coefficients. In both cases, having in mind
that
$$
\lim{}_{n\to\infty}\,\biggl|\frac{c_{n+2}x^{n+2}}{c_nx^n}\biggr|
   = \lim\nolimits_{n\to \infty}\,\biggl|\,
   \frac{\La\,n\,(n +1) - G_\mu(2 n + 1) + 2 
\nu}{(n+2)(n+1)}\,\biggr|\,\bigl|\,x^2\bigr| =
|\,\La\,|\,\bigl|\,x^2\bigr| \,,
$$
the radius of convergence $R$ is given by
$$
  R = \frac{1}{\sqrt{\,|\,\La\,|\,}} \,.
$$
Hence, when we consider the limit $\La\to0$, we recover the radius
$R=\infty$ of the Hermite's equation.

The general solution is given by the linear combination  $q = c_0
q_1 + c_1 q_2$ where $q_1(y)$ and  $q_2(y)$ are the solutions
determined by $q_1(0)=1,\ q_1'(0)=0$ and $q_2(0)=0,\ q_2'(0)=1$,
respectively. In the very particular case of the coefficient $\nu$
be given by $\nu=\nu_n$ with
$$
  2 \nu_n = (2 n + 1)G_\mu - n (n + 1)\,\La \,,
\qquad {(n\ \rm \ is\ an\ integer\ number)},
$$
then we have $c_n {\ne}0$, $c_{n+2} = 0$,
and one of the two solutions becomes a polynomial of order $n$.

The polynomial solutions are given by
\begin{itemize}
\item{} Even index (even power polynomials)
\begin{eqnarray}
  {\cal P}_{2p} &=& \sum\nolimits_{r=0}^{r=p} \,c_{2r}\,y^{2r} \cr
  c_{2r} &=& (-1)^r\,\frac{a_0}{2r\,!}\,p'\,(p'-2)(p'-4)\dots(p'-2(r-1)) \cr
     &&{\hskip20pt}\bigl[\,2G - \La(p'+1)\,\bigr]\bigl[\,2G_\mu - \La 
(p'+3)\,\bigr]
     \bigl[\,2G_\mu - \La(p'+5)\,\bigr]\dots\bigl[\,2G_\mu - 
\La(p'+2r-1)\,\bigr] \,,
\nonumber \end{eqnarray}
where we have introduced the notation $p'=2p$. More specifically,
the expressions of the first solution $q_1(y)$, in the particular
cases of $p'=0,2,4,6$, are given by:
\begin{eqnarray}
  {\cal P}_0 &=&  1  \,,\cr
  {\cal P}_2 &=&  1 - (2G_\mu - 3\La) y^2  \,,\cr
  {\cal P}_4 &=&  1 - 2(2G_\mu - 5\La) y^2 +
  (\frac{1}{3})(2G_\mu - 5\La)(2G_\mu - 7\La) y^4  \,.
\nonumber \end{eqnarray}

\item{} Odd index  (odd power polynomials)
\begin{eqnarray}
  {\cal P}_{2p+1} &=&  \sum_{r=0}^{r=p} \,a_{2r+1}\,y^{2r+1} \cr
  c_{2r+1} &=& (-1)^{r+1}\,\frac{a_1}{2r+1\,!}\,(p'-1)(p'-3)\dots(p'-(2r-1)) \cr
      &&{\hskip80pt}\bigl[\,2G_\mu - \La\,(p'+2)\,\bigr]\bigl[\,2G_\mu 
- \La (p'+4)\,\bigr]
         \dots\bigl[\,2G_\mu - \La\,(p'+2r)\,\bigr]  \,,
\nonumber \end{eqnarray}
where we have introduced the notation $p'=2p+1$. More
specifically, the expressions of the second solution $q_2(y)$ for
$p'=1,3,5$, are given by:
\begin{eqnarray}
  {\cal P}_1 &=&  y \,,\cr
  {\cal P}_3 &=&  y - (\frac{1}{3})(2G_\mu - 5\La)\,y^3 \,,\cr
  {\cal P}_5 &=&  y - (\frac{2}{3})(2G_\mu - 7\La)\,y^3
        + (\frac{1}{15})(2G_\mu - 7\La)(2G_\mu - 9\La)\,y^5  \,.\nonumber
\nonumber \end{eqnarray}
\end{itemize}

\section{Eigenfunctions $\Psi_{m,n}$ and energies $E_{m,n}$}
\subsection{Sturm-Liouville problems and $\La$-dependent Hermite
polynomials}

  We have obtained two different $\La$-dependent deformations of the
Hermite equation as well as the general solution and
the particular polynomial solutions. Nevertheless in quantum
mechanics the important point is not the equation by itself but
the associate Sturm-Liouville problem. In this case we have two
problems, one for the $Z$-equation and other for the $Y$-equation.
Moreover each one of them splits into two: one for $\La<0$
(spherical case) and other for $\La>0$ (hyperbolic case).

Spherical $\La<0$ case
  The first $\La$-dependent differential equation
\begin{equation}
   (1 + \La\,z_x^2)\, p'' + (\La - 2)\,z_x\, p' + (2 \mu -1)\, p = 0
\label{SLpz}\end{equation}
can be reduced to self-adjoint form by
making use of the following integrating factor
$$
   \mu(z_x) = (1 + \La\,z_x^2)^{-(\La +2)/(2\,\La)}  \,,
$$
in such a way that we arrive to the following expression
\begin{equation}
  \frac{d}{dz_x}\Bigl[\,A(z_x,\La)\,\frac{dp}{dz_x}\,\Bigr]
  + (2 \mu -1)\,r(z_x,\La)\,p = 0 \,,
\label{SLpz}\end{equation}
where the two functions $A=A(z_x,\La)$ and $r=r(z_x,\La)$ are given by
$$
  A = \frac{\sqrt{1 + \La\,z_x^2}}{(1 + \La\,z_x^2)^{1/\La}}
\,,\qquad
  r = \frac{1}{\sqrt{1 + \La\,z_x^2}\,(1 + \La\,z_x^2)^{1/\La}}  \,,
$$
that, together with appropriate boundary conditions, constitute a
Sturm-Liouville problem.

  If $\La$ is negative the problem is defined in the bounded interval
$[-\,a_\La,a_\La]$ with $a_\La=1/\sqrt{|\La|}$. The function
$A(z_x,\La)$ vanishes in the two end points $z_1=-\,a_\La$ and
$z_2=a_\La$ and the problem is singular. The boundary conditions
prescribe that the solutions must be bounded functions at the two
end points of the interval. The eigenvalues are the quantized values
of the parameter $\mu$, i.e. $\mu_m$, $m=0,1,2,\dots$, and the
eigenfunctions the associated polynomial solutions.

   If $\La$ is positive the problem is singular since is defined
in the whole real line $\IR$. The solutions must decrease when
$z_x\to\pm\,\infty$ in such a way that their norms, determined
with respect to the weight function $r(z_x)$, be finite. Therefore
the eigenfunctions are again the $\La$-dependent polynomials ${\cal
P}_m$, $m=0,1,2,\dots$

\begin{proposicion}
The eigenfunctions ${\cal P}_m(z_x,\La)$, $m=0,1,2,\dots$ of the de
Sturm-Liouville problem of the Eq. (\ref{SLpz})  are orthogonal
with respect to the function $r = (1+\La\,z_x^2)^{-(1/2+1/\La)}$.
\end{proposicion}
{\it Proof:}  This statement is just a consequence of the
properties of the Sturm-Liouville problems.
Because of this the polynomial solutions ${\cal P}_m$,
$m=0,1,2,\dots$, of the equation (\ref{SLpz}), satisfy
$$
  \int_{-\,a_\La}^{a_\La} \frac{{\cal P}_m(z_x,\La)\,{\cal P}_n(z_x,\La)}
  {(1 + \La\,z_x^2)^{1/\La}\,\sqrt{1 + \La\,z_x^2}} \,\, dz_x = 0
  \,,\quad m\,\ne\,n \,,\quad \La<0\,,
$$
and
$$
  \int_{-\infty}^{\infty} \frac{{\cal P}_m(z_x,\La)\,{\cal P}_n(z_x,\La)}
  {(1 + \La\,z_x^2)^{1/\La}\,\sqrt{1 + \La\,z_x^2}} \,\, dz_x = 0
  \,,\quad m\,\ne\,n \,,\quad \La>0\,.
$$
In the $\La>0$ case, as the integral is defined on a infinite
interval, the following property must be satisfied
$$
  \lim_{|z_x|\to\infty}\, z_x \,[{\cal P}_m(z_x,\La)]^2\,(1 +
  \La\,z_x^2)^{-(1/\La+1/2)} = 0  \,.
$$
The consequence is that if $\La>0$ then the quantum number $m$ is 
limited by the
condition   $m < 1/\La$, and there is only $M_\La$ eigenvalues and 
eigenfunctions
where $M_\La$ denotes the greatest integer lower than $1/\La$.
{\hfill\bb}

The following ``Rodrigues formula''
\begin{equation}
  {\cal H}_m(z_x,\La) = (-1)^n\,W_z^{1/\La+1/2}\,
  \frac{d^m}{dz_x^m}\,\Bigl[\,W_z^{m}\,W_z^{-\,(1/\La+1/2)}\,\Bigr]
  \,,\quad W_z = 1 + \La\,z_x^2  \,,
\end{equation}
leads to a family of $\La$-dependent Hermite polynomials ${\cal
H}_m$ which are proportional to ${\cal P}_m$
$$
  {\cal H}_m = k_m\,{\cal P}_m(z_x,\La) \,,\quad  m = 0,1,2,\dots
$$
where $k_m$ are constants that in the first cases are given by
\begin{eqnarray}
  k_0 &=& 1  \,,\hskip110pt
  k_1 = (2 - \La)  \,,\cr
  k_2 &=& -\,(2 - 3\La)  \,,\hskip65pt
  k_3 = -\,3(2 - 3\La)(2 - 5\La)  \,,\cr
  k_4 &=& 3(2 - 5\La)(2 - 7\La) \,,\hskip30pt
  k_5 = 15(2 - 5\La)(2 - 7\La)(2 - 9\La) \,.\nonumber
\end{eqnarray}
Alternatively we can obtain these polynomials by using the
following  function
\begin{equation}
  {\cal F}(z_x,t,\La) = \Bigl(1 + \La\,(2 t z_x - t^2)\Bigr)^{1/\La}
\end{equation}
as a generating function
\begin{equation}
  \Bigl(1 + \La\,(2 t z_x - t^2)\,\Bigr)^{1/\La}
   = \sum_{m=0}^{\infty} (\frac{1}{m\,!})\,\wt{\cal H}_m(z_x,\La)\,t^m \,,
\end{equation}
where we have used the notation $\wt{\cal H}_m$ for the
coefficients of the Taylor series. We obtain
$$
  \wt{\cal H}_m = g_m\,{\cal P}_m(z_x,\La) \,,\quad  m = 0,1,2,\dots
$$
where $g_m$ are constants that in the first cases are given by
\begin{eqnarray}
  g_0 &=& 1  \,,\hskip90pt
  g_1 = 2  \,,\cr
  g_2 &=& -\,2    \,,\hskip80pt
  g_3 = -\,12 (1 - \La)  \,,\cr
  g_4 &=& 12 (1 - \La) \,,\hskip50pt
  g_5 = 120(1 - \La)(1 - 2\La) \,.\nonumber
\end{eqnarray}

We can define the {\sl $\La$-dependent Hermite functions\/} $Z_m$
by
$$
  Z_m(z_x,\La) = {\cal H}_m(z_x,\La)\,(1 + \La\,z_x^2)^{-\,1/(2\La)}
  \,,\quad  m=0,1,2,\dots
$$
then the above statement admits the following alternative form:
{\sl The $\La$-dependent Hermite functions $Z_m(z_x,\La)$ are
orthogonal with respect to the weight function  $\wt{r}=1/\sqrt{1
+ \La\,z_x^2}$}:
$$
  \int_{-\,a_\La}^{a_\La}Z_m(z_x,\La)\,Z_n(z_x,\La)\,\wt{r}(z_x,\La) \,dz_x
  = \int_{-\,a_\La}^{a_\La}Z_m(z_x,\La)\,Z_n(z_x,\La)\frac{dz_x}
  {\sqrt{1 + \La\,z_x^2}} = 0
  \,,\quad m\,\ne\,n \,,\quad \La<0\,,
$$
and
$$
  \int_{-\infty}^{\infty}Z_m(z_x,\La)\,Z_n(z_x,\La)\,\wt{r}(z_x,\La) \,dz_x
  = \int_{-\infty}^{\infty}Z_m(z_x,\La)\,Z_n(z_x,\La)\frac{dz_x}
  {\sqrt{1 + \La\,z_x^2}} = 0
  \,,\quad m\,\ne\,n \,,\quad \La>0\,.
$$
Figures IV and V show the form of the function $Z_2(z_x,\La)$
for several values of $\La$ ($\La<0$ in Figure IV and $\La>0$ in
Figure V).

Summarizing, the final solution of the Sturm-Liouville problem for
the function $Z(z_x)$  is:
\begin{itemize}
\item{} Spherical $\La<0$ case:
\begin{eqnarray}
  &&Z_m(z_x,\La) = {\cal H}_m(z_x,\La)\,
(1 - \left|\La\right|\,z_x^2)^{\,1/(2\left|\La\right|)}\,, \cr
   &&\mu_m = \bigl(m+\frac{1}{2}\bigr) + \frac{1}{2}\,m^2\,\left|\La\right|\,,
\qquad  m = 0,1,2,\dots,m, \dots
\nonumber\end{eqnarray}
\item{} Hyperbolic $\La>0$ case:
\begin{eqnarray}
  &&Z_m(z_x,\La) = {\cal H}_m(z_x,\La)\,(1 +\La\,z_x^2)^{-\,1/(2\La)}\,, \cr
  &&\mu_m = \bigl(m+\frac{1}{2}\bigr) - \frac{1}{2}\,m^2\,\La\,,
\qquad   m= 0,1,2,\dots,M_\La.
  \nonumber\end{eqnarray}
\end{itemize}

The second $\La$-dependent differential equation
\begin{eqnarray}
  &&a_0\, q'' + a_1\, q' + a_2\, q = 0 \,,\quad
  G_m = 1 - m \La  \,,\cr
  &&a_0 = 1 + \La\,y^2 \,,\quad a_1 = 2 (\La - G_m)\,y
  \,,\quad a_2 = 2 \nu - G_m \,,\quad
\end{eqnarray} is not self-adjoint since $a'_0 \ne a_1$ but it can
be reduced to self-adjoint form by making use of the following
integrating factor
$$
  \mu(y) = (\frac{1}{a_0})\,e^{{\int}(a_1/a_0)\,dy}
  = (1 + \La\,y^2)^{-\,G_m/\La}  \,,
$$
in such a way that we arrive to the following expression
\begin{equation}
  \frac{d}{dy}\Bigl[\,B(y,m,\La)\,\frac{dq}{dy}\,\Bigr]
  + (2 \nu - G_m)\,r(y,m,\La)\,q = 0 \,,
\label{SLqy}\end{equation}
where the two functions $B=B(y,m,\La)$ and $r=r(y,m,\La)$ are given by
$$
  B = (1 + \La\,y^2)^{1 - G_m/\La} \,,\quad
  r =  (1 + \La\,y^2)^{- G_m/\La} \,.
$$
that, together with appropriate conditions for the behaviour of
the solutions at the end points, constitute a  Sturm-Liouville
problem. It is to be pointed out that the boundary conditions are
in fact different according to the sign of $\La$; therefore we
arrive to, no just one, but two different Sturm-Liouville
problems:

  If $\La$ is negative the range of the  variable $y$ is
limited by the restriction $y^2<1/|\La|$. In this case the
problem,  defined in the bounded interval
$[-\,a_\La,a_\La]$ with $a_\La=1/\sqrt{|\La|}$, is singular because
the function $q(y,\La)$ vanishes in the two end points
$y_1=-\,a_\La$ and $y_2=a_\La$. The conditions to be imposed in
this case lead to prescribe that the solutions $q(y,\La)$ of the
problem must be bounded functions at the two end points,
$y_1=-\,a_\La$ and $y_2=a_\La$, of the interval.
It is clear that this leads to the above mentioned polynomial
solutions.

  If $\La$ is positive  the variable $y$ is
defined in the whole real line $\IR$ and, therefore, the
Sturm-Liouville problem is singular. The solutions $q(y,\La)$ must
be well defined in all $\IR$,
and the boundary conditions prescribe that the behaviour of these
functions when $y\to\pm\,\infty$ must be such that their norms,
determined with respect to the weight function $r(y)$, be finite.
It is clear that in this case the solutions of the problem
are again the $\La$-dependent polynomials ${\cal H}_m$, $m=0,1,2,\dots$

\begin{proposicion}
The eigenfunctions ${\cal P}_n(y,m,\La)$, $n=0,1,2,\dots$ of the de
Sturm-Liouville problem of the Eq. (\ref{SLqy})  are orthogonal
with respect to the function $r= (1 + \La\,y^2)^{-\,G_m/\La}$,
$G_m = 1 - m \La$.
\end{proposicion}
{\it Proof:} This statement is just a consequence of the
properties of the Sturm-Liouville problems.
Because of this the polynomial solutions ${\cal P}_n={\cal P}_n(y,m,\La)$,
$n=0,1,2,\dots$, of the equation (\ref{SLqy}), satisfy
$$
  \int_{-\,a_\La}^{a_\La} \frac{{\cal P}_r(y,m,\La)\,{\cal P}_s(y,m,\La)}
  {(1 + \La\,y^2)^{G_m/\La}} \,\, dy = 0
  \,,\quad r\,\ne\,s \,,\quad \La<0\,,
$$
and
$$
  \int_{-\infty}^{\infty} \frac{{\cal P}_r(y,m,\La)\,{\cal P}_s(y,m,\La)}
  {(1 + \La\,y^2)^{G_m/\La}} \,\, dy = 0
   \,,\quad r\,\ne\,s \,,\quad \La>0\,.
$$
In the $\La>0$ case, as the integral is defined on a infinite
interval, the following property must be satisfied
$$
  \lim_{|y|\to\infty}\, y \,[{\cal P}_n(y,m,\La)]^2\,(1 +
  \La\,y^2)^{m-1/\La} = 0  \,,
$$
and as a consequence, the quantum number $n$ is limited by the
condition
$$
  n < \frac{1}{\La} - m - \frac{1}{2} \,.
$$
That is, for every value of $m$ there is only $N_\La$ eigenvalues and 
eigenfunctions
where $N_\La$ denotes the greatest integer number lower than $1/\La-m-1/2$.
{\hfill\bb}

The ``Rodrigues formula'' for these new family of $\La$-dependent
Hermite polynomials is given by
\begin{equation}
  {\cal H}_n(y,m,\La) = (-1)^n\,W_y^{G_m/\La}\,
  \frac{d^n}{dy^n}\,\Bigl[\,W_y^{n}\,W_y^{-\,G_m/\La}\,\Bigr]
  \,,\quad W_y = 1 + \La\,y^2  \,,
\end{equation}
in such a way that we obtain
$$
  {\cal H}_n = k_n\,{\cal P}_n(y,m,\La) \,,\quad  n = 0,1,2,\dots
$$
with the following values for the first constants
\begin{eqnarray}
  k_0 &=& 1  \,,\hskip135pt
  k_1 = 2 (G_m - \La)  \,,\cr
  k_2 &=& -\,2 (G_m - 2\La)   \,,\hskip75pt
  k_3 = -\,12(G_m - 2\La)(G_m - 3\La)  \,,\cr
  k_4 &=& 12(G_m - 2\La)(G_m - 3\La) \,,\hskip30pt
  k_5 = 120(G_m - 3\La)(G_m - 4\La)(G_m - 5\La) \,.\nonumber
\end{eqnarray}
The $(\La,m)$-dependent function defined by
\begin{equation}
  {\cal F}(y,t,\La) = \Bigl(1 + \La\,(2 t y - t^2)\Bigr)^{(G_m/\La-1/2)}
\end{equation}
is a generating function with the following power expansion
\begin{equation}
  \Bigl(1 + \La\,(2 t y - t^2)\,\Bigr)^{(G_m/\La-1/2)}
   = \sum_{m=0}^{\infty} \bigl(\frac{1}{n\,!}\bigr)\,\wt{\cal H}_n(y,m,\La)\,t^n
\end{equation}
where we have used the notation $\wt{\caH}_n$ for the coefficients
of the Taylor series. The first $(\La,m)$-dependent Hermite
polynomials obtained in such a way have the following expressions
$$
  \wt{\cal H}_n = g_n\,{\cal P}_n(y,m,\La) \,,\quad  n = 0,1,2,\dots
$$
where the constants $g_i$, $i=0,1,2,\dots,5$, take the values
\begin{eqnarray}
  g_0 &=& 1  \,,\hskip140pt
  g_1 =  (2G_m - \La)  \,,\cr
  g_2 &=&  -\,(2G_m - \La)   \,,\hskip80pt
  g_3 =  -\, 3  (2G_m - \La) (2G_m - 3\La)  \,,\cr
  g_4 &=& 3 (2G_m- \La) (2G_m - 3\La) \,,\hskip30pt
  g_5 =  15 (2G_m - \La) (2G_m - 3\La) (2G_m - 5\La)  \,.\nonumber
\end{eqnarray}

The {\sl $\La$-dependent Hermite functions} $Y_n$ are defined by
$$
  Y_n(y,m,\La) = {\cal H}_n(y,m,\La)\,(1 + \La\,y^2)^{-\,G_m/(2\La)}
  \,,\quad  n=0,1,2,\dots
$$
and the above statement admits the following alternative form:
{\sl The $\La$-dependent Hermite functions $Y_n(y,m,\La)$ are
orthogonal with respect to the weight function $\wt{r}=1$}:
$$
  \int_{-\,a_\La}^{a_\La}Y_r(y,m,\La)\,Y_s(y,m,\La)\,\wt{r}(y,\La) \,dy
  = \int_{-\,a_\La}^{a_\La}Y_r(y,m,\La)\,Y_s(y,m,\La)\, dy = 0
  \,,\quad r\,\ne\,s \,,\quad \La<0\,,
$$
and
$$
  \int_{-\infty}^{\infty}Y_r(y,m,\La)\,Y_s(y,m,\La)\,\wt{r}(y,\La) \,dy
  = \int_{-\infty}^{\infty}Y_r(y,m,\La)\,Y_s(y,m,\La)\, dy = 0
  \,,\quad r\,\ne\,s \,,\quad \La>0\,.
$$

Summarizing, the final solution of the Sturm-Liouville problem for
the function $Y(y)$  is:
\begin{itemize}
\item{} Spherical $\La<0$ case:
{\openup3pt\begin{eqnarray}
  &&Y_n(y,m,\La) = {\cal H}_n(y,m,\La)\,
  (1 - \left|\La\right|\,y^2)^{\,G_m/(2\left|\La\right|)}\,, \quad
   G_m = 1 + m \,|\La|  \,, \cr
   &&\nu_n = \bigl(n+\frac{1}{2}\bigr)\,G_m + \frac{1}{2}  \,n\,(n+1)\,|\La|\,,
\qquad  n = 0,1,2,\dots,m, \dots
\nonumber\end{eqnarray}}
\item{} Hyperbolic $\La>0$ case:
{\openup3pt\begin{eqnarray}
  &&Y_n(y,m,\La) = {\cal H}_n(y,m,\La)\,(1 +\La\,y^2)^{-\,G_m/(2\La)}\,,
  \quad  G_m = 1 - m \La    \,,\cr
  &&\nu_n =\bigl(n+\frac{1}{2}\bigr)\,G_m - \frac{1}{2}  \,n\,(n+1)\,\La\,,
\qquad   n= 0,1,2,\dots,N_\La.
  \nonumber\end{eqnarray}}
\end{itemize}

\subsection{Final solution}

  The wave functions of the $\La$-dependent nonlinear oscillator are
$$
  \Psi_{m,n}(z_x,y)  = Z_m(z_x)\,Y_n(y)  \,,\quad
  z_x = \frac{x}{\sqrt{\,1 + \La\,y^2\,}} \,,
$$
with energies given by
$$
  e_{m,n} = \mu_m + \nu_n = \Bigl((m+\frac{1}{2}) + (n+\frac{1}{2})\Bigr)
\Bigl[\, 1 - \frac{1}{2}(m+n)\La\,\Bigr]  \,.
$$
So the total energy $E_{m,n}= (\hbar\,\al) e_{m,n}$ is a linear
function of $\La$ and depends, as in the $\La=0$ case, of the sum
$m+n$ of the two quantum numbers.
It is clear that $\Psi_{m,n}$ is well defined for any value of $\La$ and that
the following limit is satisfied
$$
  \lim{}_{\La\to 0} \Psi_{m,n}(z_x,y) = H_m(x) H_n(y)
\,e^{-\,(1/2)\,(x^2+y^2)}
  \,,\quad m,n=0,1,2,\dots
$$

  We recall that this approach has considered the space $L^2(\IR^2,d\mu_\La)$
as the appropriate Hilbert space.
Thus, if $\Psi_{m,n}(x,y)$ and $\Psi_{r,s}(x,y)$ are to wave functions
representing states of the nonlinear oscillators with quantum
number $(m,n)$ and $(r,s)$ respectively, then the scalar product
is given by
$$
  \langle \Psi_{m,n}\,,\Psi_{r,s}\rangle_{\La}
  = \int \Psi_{m,n}(x,y)\,\Psi_{r,s}(x,y) \,d\mu_\La \,.
$$
The point is that  making use of the equality
$$
  1+\La\,r^2 = (1+\La\,z_x^2)\,(1+\La\,y^2)
$$
the measure  $d\mu_\La$ becomes as follows
$$
  d\mu_\La = \Bigl(\frac{1}{\sqrt{1+\La\,r^2}}\Bigr)\,dx\,dy =
  \Bigl(\frac{1}{\sqrt{1+\La\,z_x^2}}\Bigr)\,dz_x\,dy
$$
in coordinates $(z_x,y)$.
Hence, making use of the factorization of the measure $d\mu_\La$ in
coordinates $(z_x,y)$, we can factorize the scalar product in
$\IR^2$ as a product of two one-dimensional scalar products
and arrive to the following important property
\begin{eqnarray}
  \langle \Psi_{m,n}\,,\Psi_{r,s}\rangle_{\La}
  &=& \int  Z_m(z_x)\,Z_r(z_x)\,Y_n(y)\,Y_s(y)\,\,d\mu_\La  \cr
  &=& \Biggl(\int Z_m(z_x,\La)\,Z_r(z_x,\La)\frac{dz_x}
  {\sqrt{1 + \La\,z_x^2}} \Biggr)
  \Biggl(\int  Y_n(y,\La)\,Y_s(y,\La)\, dy\Biggr)  \cr
  &=&   \delta_{m,r}\, \delta_{n,s}  \,.
\nonumber\end{eqnarray}}

The following two points summarize the main characteristics of the
energies of the bound states:

\begin{enumerate}

\item{} Spherical $\La<0$ case:

  The Hamiltonian $\widehat{H}(\La)$ describes a quantum oscillator
on the sphere $S_{\k}^2$ ($\k>0$).
The oscillator possesses a countable infinite set of bound states
$\Psi_{n,m}$, with $n,m=0,1,2,\dots$ and the energy spectrum is
unbounded, not equidistant and with a difference between the
levels that increases with $N$
\begin{eqnarray}
  &&e_0<e_1<e_2<e_3<\dots<e_N<e_{N+1}<\dots \cr
  &&e_{N+1} - e_N = 1 + \left(N+1\right)\,\left|\La\right| \,,\quad
  N = m + n\,.
\nonumber\end{eqnarray}
The oscillations of the wave functions are reinforced and the values
of the energies $E_{n,m}$ are higher than in the Euclidean $\La=0$ case;
that is, $E_{n,m}(\La)>E_{n,m}(0)$.

\item{} Hyperbolic $\La>0$ case:

The Hamiltonian $\widehat{H}(\La)$ describes a
quantum oscillator on the hyperbolic plane $H_{\k}^2$ ($\k<0$).
The oscillator possesses only a finite number of bound states
$\Psi_{n,m}$,  with $n+m=0,1,2,\dots, N_\La$, $N_\La<1/\La-1/2$, and the
energy spectrum is bounded, not equidistant and with a difference
between the levels that decreases with $N$
\begin{eqnarray}
  &&e_0<e_1<e_2<e_3<\dots<e_{N_\La} \cr
  &&e_{N+1} - e_N = 1 - \left(N+1\right)\,\La \,,\quad
  N = m + n\,.
\nonumber\end{eqnarray}
The oscillations of the wave functions are smoothed down and the
values of the energies $E_{n,m}$ is lower than in the Euclidean
$\La=0$ case; that is $E_{n,m}(\La)<E_{n,m}(0)$.

\end{enumerate}

The degeneracy of the energy levels is the same that in the Euclidean case.

Figures VI and VII show the values of the energy $e_{m,n}$ as a
function of $N$, $N=m+n$, for several values of $\La$.
Figure VI shows as $E_{n,m}(\La)<E_{n,m}(0)$ when $\La>0$
(hyperbolic $\La>0$ case)
and $E_{n,m}(\La)>E_{n,m}(0)$ when $\La<0$
(spherical $\La<0$ case).
Figure VII shows the plot of the energy $E_{m,n}$ for three different
values of $\La>0$; it is clear that when $\La$ decreases the maximum
of the curve moves into the up right and the number of bound sates
increases.

\section{Resolution of the Schr\"odinger equation II}

The second alternative way of solving the Schr\"odinger equation 
(\ref{EcSchxy})
is using the property of separability in coordinates $(x,z_y)$ with $z_y$
defined by $z_y = y/\sqrt{1 + \La\,x^2}$\,.
This second approach is symmetric to the first one so the solution
can be directly given as
$$
  \Phi_{n,m}(x,z_y)  = X_n(x)\,Z_m(z_y)  \,,
$$
with
\begin{eqnarray}
&& X_n(x,m,\La) = {\cal H}_n(x,m,\La)\,(1 +\La\,x^2)^{-\,(1 - m
\La)/(2\La)}\,,\cr
&& Z_m(z_y,\La) = {\cal H}_m(z_y,\La)\,(1
+\La\,z_y^2)^{-\,1/(2\La)}\,. \nonumber
\end{eqnarray} The
interesting property of this solution is that it satisfies
\begin{eqnarray}
&&\bigl(\widehat{H_1} - \la\, \widehat{J}^2\bigr)\,\Phi_{n,m}(x,z_y) =
\nu_{n}\,\Phi_{n,m}(x,z_y)  \,,\cr
&&\widehat{H_2}\,\Phi_{n,m}(x,z_y) = \mu_m\,\Phi_{n,m}(x,z_y) \,.
\nonumber\end{eqnarray}

  It is interesting to relate the existence of this second
alternative approach with the above mentioned property of
superintegrability.

There is not presently a satisfactory definition of quantum integrability.
The most direct way of considering this question is by direct translation
of the classical notions, so that, according to this approach, a quantum
Hamiltonian $\widehat{H}$ would be integrable if
there exists a set $\{A_i\}$ of $n$ independent observables
(including the Hamiltonian itself) that  pairwise commute. If
there exists an additional set of independent observables
$\{B_r\}$ commuting with $\widehat{H}$ then the system would be
superintegrable \cite{LeVi95}-\cite{CaNeOl06}; of course the
$\{B_r\}$ do not necessary commute between them and every $B_r$
only commute with some of the $\{A_i\}$. The main problem is the
definition of independence for quantum operators since the
commutation relation $[A_j,A_k]=0$ can be considered as determining a 
functional
dependence between $A_j$ and $A_k$. Several criteria have been
analyzed; the most simple is to consider the operators $\{A_i\}$
as independent if they are obtained by quantizating classical functions which
are functionally independent.

On the other hand most of classical superintegrable systems are actually
superseparable (they admit separations of variables in at least
two coordinate systems) and the classical integrals of motion are at most
quadratic in the momenta; in this case all the operators $\{A_i, B_r\}$ are
first or second order differential operators and
this restriction makes easier the study of independence (when
third order operators are considered then quantum
integrability can lead to properties rather different to those
of the classical system \cite{Hi98,RoWi02}).
Let us mention that although quantum
superintegrability and exact solvability are defined in different
ways, it has been conjectured \cite{RoWi02} that all maximally
superintegrable systems are exactly solvable.

  A consequence of the quantum superintegrability is that, as we have
a total of $2n-1$ operators commuting with $\widehat{H}$, we can
construct different complete sets of $n$ commuting observables
and, therefore, different ways of characterizing the wave function
$\Psi$. Concerning this quantum $\la$-dependent oscillator, it is
endowed with the following three sets of commuting observables
$$
  \{ \widehat{H_1}\,,\widehat{H_2} - \la\, \widehat{J}^2 \}\,,{\hskip 10pt}
  \{ \widehat{H_1} - \la\, \widehat{J}^2\,,\widehat{H_2} \}\,,{\hskip 10pt}
  \{ \widehat{H_1} + \widehat{H_2}\,, \widehat{J}^2 \} \,,
$$
that correspond to three different ways of decomposing
$\widehat{H}$ as a sum of two commuting observables and also to
three coordinate systems separating the Schr\"odinger equation
$$
  (z_x,y) \,,{\hskip 10pt} (x,z_y) \,,{\hskip 10pt} (r,\phi) \,,
$$
and to three alternative ways of representing the wave function.

\section{Final comments and outlook}

   The harmonic oscillator is not a specific or special characteristic
of the Euclidean space but it is well defined in all the three
spaces of constant curvature. In fact if we use the curvature $\k$
(or $\la$) as a parameter then we can say that there are not three
different harmonic oscillators but only one that is defined, at
the same time, in the three manifolds. This property, that was
known at the classical level, is also true for the quantum system.

  If we consider the spherical and hyperbolic systems as a
deformation of the well known Euclidean system (in the sense
discussed in the Introduction) then this deformation appears as
clearly asymmetric. This is a natural result since the sphere
$S^2$ and the hyperbolic plane $H^2$ are geometrically different
and, because of this, some dynamical properties, as the
characteristics of the wave functions $\Psi_{m,n}$ and the
energies $E_{m,n}$, also show differences depending of the sign of
$\la$.

  We finalize with some open questions.

Firstly, we have focussed our study on the systems $(x,z_y)$ and
$(z_x,y)$ because of their relations with the Cartesian
coordinates and the Hermite polynomials in the Euclidean $\la=0$
case. Nevertheless the resolution in polar coordinate must also be
studied.

Secondly, we have quantized the system by analyzing the symmetries
of the metric, obtaining an invariant measure and expressing the
Hamiltonian as a function of the Noether momenta. The use of this
quantization procedure for other systems with a position-dependent
mass is a matter to be studied.

Thirdly, one of the byproducts of this study is the existence of
$\la$-dependent deformations (or generalizations) of the Hermite
polynomials endowed with the appropriate  properties of
orthogonality, Rodrigues formula and generating functions. The
first family was already obtained in Ref. \cite{CRS06AnPh} but the
second family is new; it is clear that the three-dimensional
oscillator will lead to a new third family.  These polynomials are
interesting and deserve a more detailed study not only for the
quantum problem but also from a mathematical viewpoint.  At this
point we also recall the existence of another family of
``relativistic Hermite polynomials" obtained by Aldaya et al in
Ref. \cite{AlBiN91, AlBiGu96, AlGu05} in the study of the
relativistic quantum harmonic oscillator. We also note that the
equations (\ref{Ecpz}) and (\ref{Ecqy}) are of hypergeometric type
and the relation of the solutions with the hypergeometric series
is an interesting problem.

Four, it was proved that the one-dimensional $\la$-oscillator is
solvable \cite{CRS06AnPh} by the use of the Schr\"odinger
factorization formalism in terms of first order differential
operators $A$ and $A^+$ (these operators are known as intertwining
operators). The factorization of two-dimensional systems still
remains as a very difficult problem but, in any case, the
existence of appropriate operators $A$ and $A^+$ for this
two-dimensional oscillator must be studied.

Finally,  the technique of introducing the curvature $\kappa$ as a
parameter for the joint analysis of the dynamics in the three
manifolds ($S_{\kappa}^2, E^2, H_{\kappa}^2$) has been generalized
to the Cayley-Klein geometries (see Refs. \cite{BaHeOlS93}--\cite{CaRSS05}).
In this more general case this technique is used
with {\it two parameters, $\kappa_1$ and $\kappa_2$}, which
correspond to a space $M_{\kappa_1,\kappa_2}$ with constant
curvature $\kappa_1$ and signature $(+1, \kappa_2$). This
formalism is more general and includes the $\kappa$-dependent
formalism as the particular case $\kappa_1=\kappa$ and
$\kappa_2=1$. We think that the study presented in this article
for the quantum harmonic oscillator on the three classical spaces
of constant curvature can be extended to the de-Sitter, Minkowski
and anti-de-Sitter spaces by using this two parameters formalism.

\section*{Appendix: Geodesic polar coordinates}

    A two--dimensional manifold $M$ can be described by using different
coordinate systems.
If we consider it as an imbedded submanifold of $\IR^3$, then the points
of $M$ can be characterized by the three external coordinates,
as $(x,y,z)$ or $(r,\phi,\te)$, plus an additional constraint relation.
Nevertheless, in differential geometric terms, a more appropriate approach
is to develop the study by using two--dimensional systems of coordinates
adapted to $M$.

    On any general two--dimensional Riemannian space,
not necessarily of constant curvature, there are two distinguished
types of local coordinate systems: ``geodesic parallel" and
``geo\-desic polar" coordinates. They reduce to the familiar
Cartesian and polar coordinates on the Euclidean plane and both
are based on a origin point $O$ and an oriented geodesic $g_1$ through $O$.

   For any point $P$ in some suitable neighbourhood of $O$, there is a
unique geodesic $g$ joining $P$ with $O$. The (geodesic) polar
coordinates $(R,\Phi)$ of $P$, relative to the origin $O$ and the
positive geodesic ray of $g_1$, are the (positive) distance $R$
between $P$ and $O$ measured along $g$, and the angle $\Phi$
between $g$ and the positive ray $g_1$, measured around $O$.
These coordinates are defined in a neighbourhood of $O$ not
extending beyond the cut locus of $O$; polar coordinates are
singular at $O$, and $\Phi$ is discontinuous on the positive ray
of $g_1$.

   In the case of $M$ being a space of constant curvature $\k$,
the expression for the differential element of distance $ds^2$
is given by
$$
  ds_\k^2 = d R^2 + \Sin_\k^2(R)\,d{\Phi}^2 \,,
$$
so that we get $ds^2 = d r^2 + r^2\,d{\phi}^2$ for the
particular $\k=0$ Euclidean case.

\section*{\bf Acknowledgments.}
Support of projects  BFM-2003-02532, FPA-2003-02948, E23/1 (DGA),
MTM-2005-09183, and VA-013C05 is acknowledged.

{\small

    }
\vfill\eject
\section*{\bf Figure Captions}

\begin{itemize}

\item{}{\sc Figure I}.{\enskip}
Plot of the one-dimensional potential $V(\la)=(1/2)\,(\al^2 x^2)/(1 + 
\la\,x^2)$,
as a function of $x$, for $\al=1$ and $\la<0$.

\item{}{\sc Figure II}.{\enskip}
Plot of the one-dimensional potential $V(\la)=(1/2)\,(\al^2 x^2)/(1 + 
\la\,x^2)$,
as a function of $x$, for $\al=1$ and $\la>0$.

\item{} {\sc Figure III}.{\enskip} Plot of the potential  $U_\k(r)$,
$\al=1$, as a function of $r$, for $\k=-1$ (lower curve), $\k=0$
(dash line), and $\k=1$ (upper curve).

\item{} {\sc Figure IV}.{\enskip} Plot of the $\La$-dependent Hermite
function $Z_2(z_x,\La)$ as a function of $z_x$ for $\La=0$ (dashed
curve) and $\La=-0.15$  and $\La=-0.30$. For very small values of
$|\La|$ the figure is very similar to the standard Hermite curve
and when the value of $|\La|$ increases the oscillations become
stronger.

\item{} {\sc Figure V}.{\enskip} Plot of the $\La$-dependent Hermite
function $Z_2(z_x,\La)$ as a function of $z_x$ for $\La=0$ (dashed
curve) and $\La=0.15$  and $\La=0.30$. In this case when the value
of $\La$ increases the oscillations become softer and for $\La\ge
0.5$ the Hermite function becomes not normalizable.

\item{} {\sc Figure VI}.{\enskip}  Plot of the energy $e_N$ as a function
of $N$, $N=m+n$, for $\La=0.30$ (lower curve) and $\La=-0.30$
(upper curve). The thick points $(N,e_N)$, corresponding to the
values $N=0, 1, 2$, represent the three bound states existing for
$\La=0.30$ and the first three bound states for $\La=-0.30$. The
straight line (dashed line) placed in the middle corresponds to
the linear harmonic oscillator.

\item{} {\sc Figure VII}.{\enskip} Plot of the energy $e_{m,n}$ as a
function of $N$, $N=m+n$, for $\La=0.45$ (lower curve), $\La=0.30$
(middle curve) and $\La=0.15$ (upper curve). The curves also show
the plot of the points $(N,e_N)$ for the values $N=0,1$, $N=0,
1,2$, and $N=0, 1, \dots, 6$, respectively. Every thick point
represents a certain number of bound states with the same energy
$E_{m,n}$ and characterized by quantum numbers $m$ and $n$ such
that $n+m=N$.  When $\La$ decreases the maximum of the curve moves
into the up right, the number of bound sates goes up and in the
limit $\La\to 0$ the curve converges into a straight line parallel
to the diagonal (dashed line).

\end{itemize}
\vfill\eject

\centerline{\epsfbox{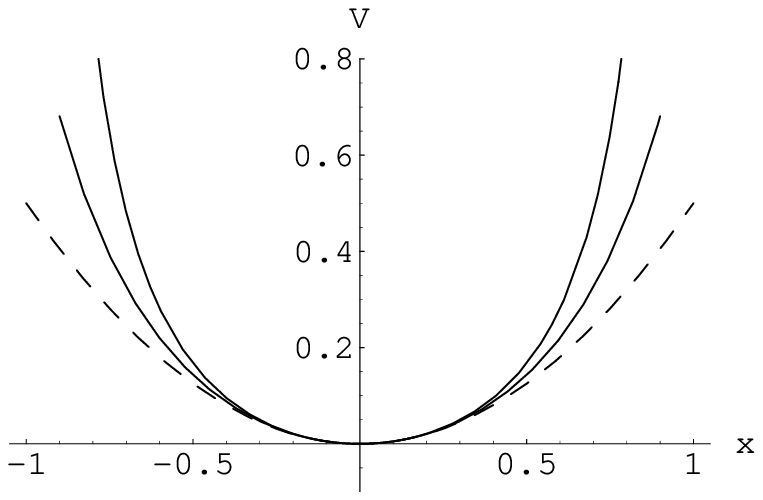} }

{\sc Figure I}.{\enskip} Plot of the one-dimensional potential
$V(\la)=(1/2)\,(\al^2 x^2)/(1 + \la\,x^2)$, as a function of $x$,
for $\al=1$ and $\la<0$.

\vskip50pt

\centerline{\epsfbox{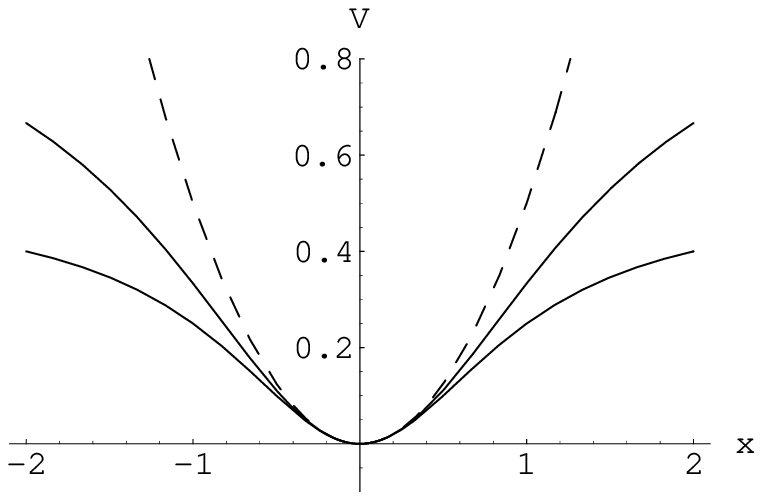} }

{\sc Figure II}.{\enskip}  Plot of the one-dimensional potential
$V(\la)=(1/2)\,(\al^2 x^2)/(1 + \la\,x^2)$, as a function of $x$,
for $\al=1$ and $\la>0$.

\newpage

\centerline{\epsfbox{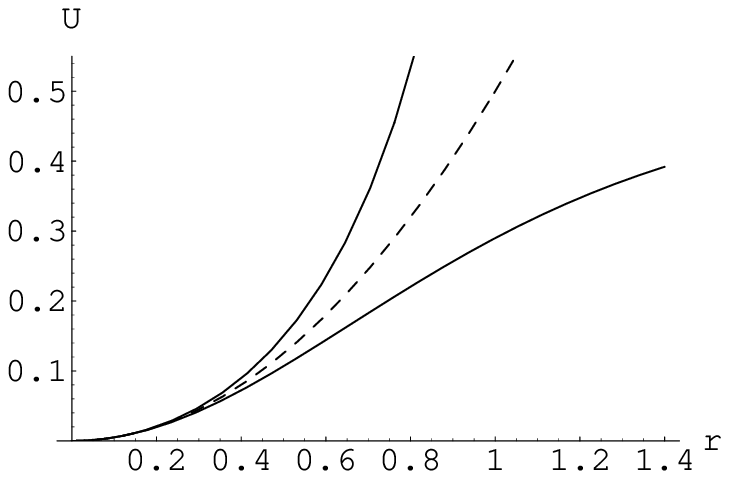} }

{\sc Figure III}.{\enskip}  Plot of the potential  $U_\k(r)$,
$\al=1$, as a function of $r$, for $\k=-1$ (lower curve), $\k=0$
(dash line), and $\k=1$ (upper curve).

\vskip50pt

\centerline{\epsfbox{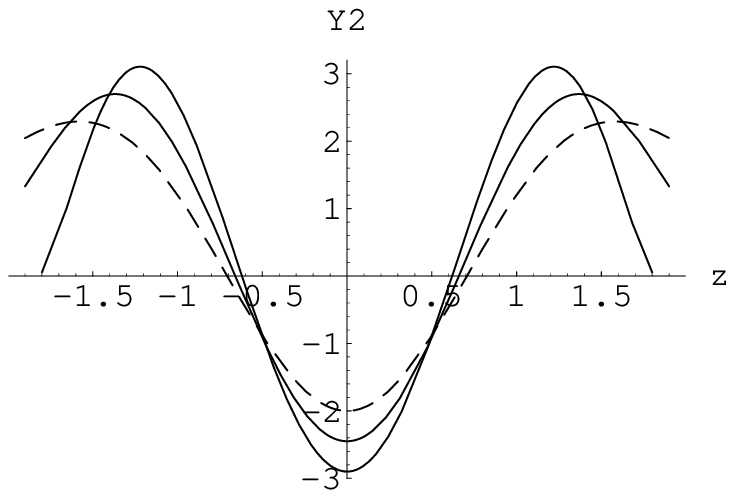} }

{\sc Figure IV}.{\enskip}  Plot of the $\La$-dependent Hermite
function $Z_2(z_x,\La)$ as a function of $z_x$ for $\La=0$ (dashed
curve) and $\La=-0.15$  and $\La=-0.30$. For very small values of
$|\La|$ the figure is very similar to the standard Hermite curve
and when the value of $|\La|$ increases the oscillations become
stronger.

\newpage

\centerline{ \epsfbox{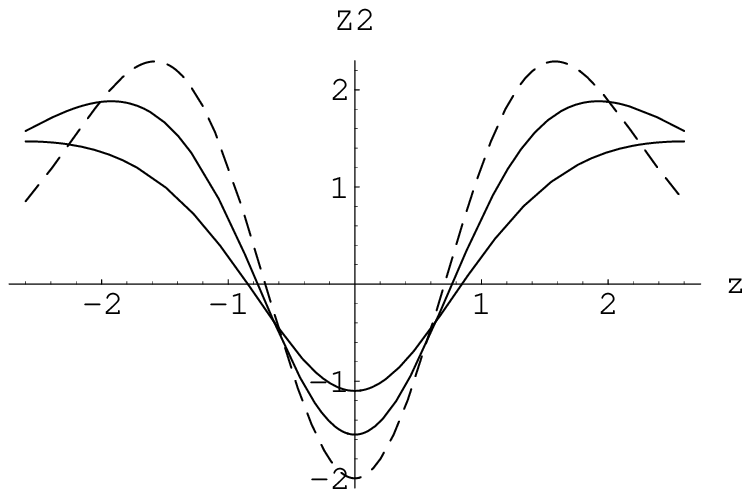} }

{\sc Figure V}.{\enskip} Plot of the $\La$-dependent Hermite
function $Z_2(z_x,\La)$ as a function of $z_x$ for $\La=0$ (dashed
curve) and $\La=0.15$  and $\La=0.30$. In this case when the value
of $\La$ increases the oscillations become softer and for $\La\ge
0.5$ the Hermite function becomes not normalizable.

\vskip50pt

\centerline{\epsfbox{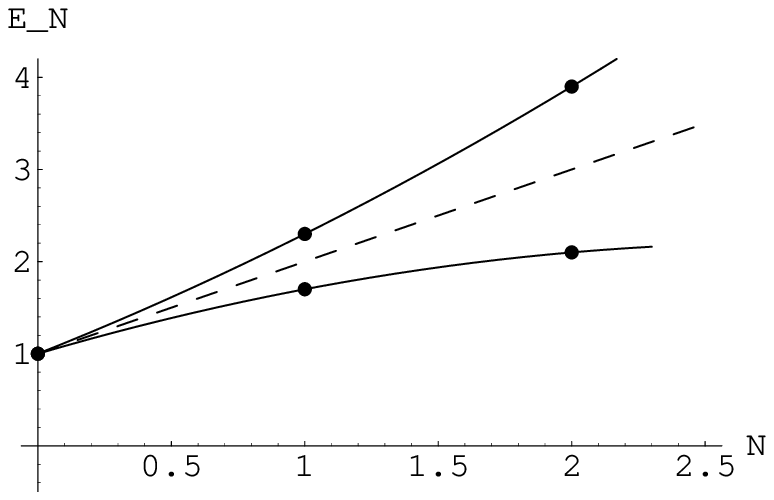} }

{\sc Figure VI}.{\enskip}  Plot of the energy $e_N$ as a function
of $N$, $N=m+n$, for $\La=0.30$ (lower curve) and $\La=-0.30$
(upper curve). The thick points $(N,e_N)$, corresponding to the
values $N=0, 1, 2$, represent the three bound states existing for
$\La=0.30$ and the first three bound states for $\La=-0.30$. The
straight line (dashed line) placed in the middle corresponds to
the linear harmonic oscillator.

\newpage

\centerline{ \epsfbox{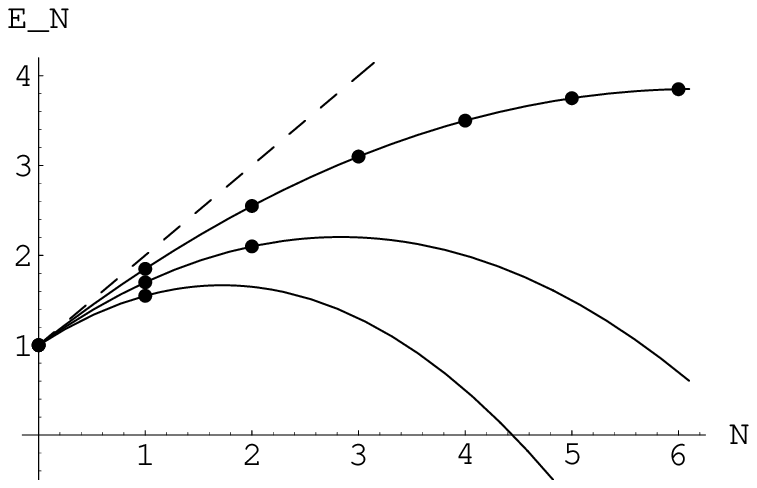} }

{\sc Figure VII}.{\enskip} Plot of the energy $E_{m,n}$ as a
function of $N$, $N=m+n$, for $\La=0.45$ (lower curve), $\La=0.30$
(middle curve) and $\La=0.15$ (upper curve). The curves also show
the plot of the points $(N,E_N)$ for the values $N=0,1$, $N=0,
1,2$, and $N=0, 1, \dots, 6$, respectively. Every thick point
represents a certain number of bound states with the same energy
$E_{m,n}$ and characterized by quantum numbers $m$ and $n$ such
that $n+m=N$.  When $\La$ decreases the maximum of the curve moves
into the up right, the number of bound sates goes up and in the
limit $\La\to 0$ the curve converges into a straight line parallel
to the diagonal (dashed line).

\end{document}